\documentclass[letterpaper,11pt]{article} 
\usepackage{arydshln}
\usepackage{amsmath}
\usepackage{amssymb}
\usepackage[dvips]{epsfig}
\usepackage{dcolumn}
\usepackage{enumerate}
\usepackage{hhline}
\usepackage{dsfont}
\usepackage{afterpage}
\usepackage{arydshln}
\usepackage{graphicx}
\usepackage{color}
\usepackage[usenames,dvipsnames,table]{xcolor}
\usepackage{colortbl}
\definecolor{Gray}{gray}{0.9}
\usepackage{rotating}
\usepackage[breaklinks]{hyperref}
\usepackage{breakurl} 
\usepackage{xr}
\usepackage[percent]{overpic}
\usepackage[]{caption}
\usepackage{algorithmicx}
\usepackage[]{algpseudocode}
\usepackage{algorithm}
\usepackage{diagbox}
\usepackage{graphicx}
\usepackage{wrapfig}
\usepackage{lscape}
\usepackage{fontenc}
\usepackage{setspace}
\usepackage{bm}
\usepackage{slashbox}
\usepackage{lscape}
\usepackage{breakurl} 
\usepackage{multirow,booktabs}
\usepackage{eurosym}
\usepackage{titlesec}
\usepackage{sectsty}
\usepackage{placeins}
\usepackage{subcaption}
\usepackage[flushleft]{threeparttable}
\usepackage{makecell}
\usepackage[normalem]{ulem}
\usepackage{tikz}
\usepackage[most]{tcolorbox}
\epsfverbosetrue
\def\theequation{\thesection.\arabic{equation}}  
\def\abstract{\if@twocolumn
\section*{Abstract}
\else \normalsize 
\begin{center}
{\bf Summary\vspace{-.5em}\vspace{0pt}} 
\end{center}
\quotation 
\fi}
\def\endabstract{\if@twocolumn\else\endquotation\fi}

\makeatletter
\newcommand{\myappendix}[1]{
	\setcounter{section}{1}
        \renewcommand{\thesection}{A\arabic{section}}}

\def \calB {\mathcal B}
\def \calC {\mathcal C}

\def \calL {\mathcal L}

\def \calR {\mathcal R}

\def \avec {\text{\boldmath$a$}}

\def \dvec {\text{\boldmath$d$}}

\def \gvec {\text{\boldmath$g$}}

\def \pvec {\text{\boldmath$p$}}

\def \uvec {\text{\boldmath$u$}}    
\def \vvec {\text{\boldmath$v$}}    
    
\def \xvec {\text{\boldmath$x$}}    
\def \yvec {\text{\boldmath$y$}}    
\def \zvec {\text{\boldmath$z$}}    

\def \zerovec {\text{\boldmath$0$}}

\def \alphavec        {\text{\boldmath$\alpha$}}
\def \betavec         {\text{\boldmath$\beta$}}

\def \epsilonvec      {\text{\boldmath$\epsilon$}}

\def \thetavec        {\text{\boldmath$\theta$}}
\def \varthetavec     {\text{\boldmath$\vartheta$}}

\def \lambdavec       {\text{\boldmath$\lambda$}}
\def \muvec           {\text{\boldmath$\mu$}}

\def \xivec           {\text{\boldmath$\xi$}}

\def \tauvec          {\text{\boldmath$\tau$}}

\def \psivec          {\text{\boldmath$\psi$}}

\usepackage{color}
\usepackage{colordvi}
\newlength{\breite}
\breite\textwidth
\newcounter{aufg}[section]
  {\refstepcounter{aufg}\noindent\textbf{Exercise \arabic{aufg}:}
   \\*[1ex]\noindent}{\vspace{.5cm}}
   
 \newcounter{notes}[section]
  {\refstepcounter{aufg}\noindent\textbf{}
   \\*[1ex]\noindent}{\vspace{.5cm}}
   
\usepackage{amsthm}  % Theoreme
\theoremstyle{definition}

\newtheorem*{beisp*}{Example}
\newtheorem{Proof}{Proof}


\newtheoremstyle{break}% name
  {}%         Space above, empty = `usual value'
  {}%         Space below
  {}% Body font
  {}%         Indent amount (empty = no indent, \parindent = para indent)
  {\bfseries}% Thm head font
  {.}%        Punctuation after thm head
  {\newline}% Space after thm head: \newline = linebreak
  {}%         Thm head spec
  
\theoremstyle{break}

\newcommand{\head}[2]%
 {\hrule \vspace{.15cm} {\sfbold Advanced Statistical Inference, Summer Term 2012, Georg-August-University G\"ottingen}\hfill
{\sfbold Sheet #1}\\
{\sfbold Prof. Dr. Thomas Kneib, Nadja Klein}\hfill {\sfbold #2}

\vspace{.2cm}
\hrule

\vspace{1cm}

}

\newcounter{auf}
{\refstepcounter{auf}
\begin{center}
\fcolorbox[gray]{0}{.95}{
\makebox[\breite]{
\textbf{Exercise \arabic{auf}}
}}\\*[1ex]\noindent
\end{center}
}{\vspace{.5cm}}

\newcounter{loes}[section]
{\stepcounter{loes}
\begin{center}
\fcolorbox[gray]{0}{.95}{
\makebox[\breite]{
\textbf{L"osung \arabic{loes}}
}}\\*[1ex]\noindent
\end{center}
}{}

{\begin{center}
\fcolorbox[gray]{0}{.95}{
\makebox[\breite]{
\textbf{Zu Aufgabe #1}
}}\\*[1ex]\noindent
\end{center}\vspace{1cm}
}{\vspace{1cm}}

\newcounter{ka}
{\refstepcounter{ka}
\begin{center}
\framebox[\textwidth]{
\textbf{Aufgabe \arabic{ka}} \hfill #1 Punkte
}\\*[1ex]\noindent
\end{center}
}{\vspace{1cm}}

\newcounter{lka}
{\refstepcounter{lka}
\begin{center}
\framebox[\textwidth]{
\textbf{L\"osung \arabic{lka}} \hfill #1 Punkte
}\\*[1ex]\noindent
\end{center}
}{\vspace{1cm}}

\usepackage[margin=1in]{geometry}
\titlespacing*\section{0pt}{0pt plus 4pt minus 2pt}{0pt plus 2pt minus 2pt}
\titlespacing*\subsection{0pt}{0pt plus 4pt minus 2pt}{0pt plus 2pt minus 2pt}
\titlespacing*\subsubsection{0pt}{0pt plus 4pt minus 2pt}{0pt plus 2pt minus 2pt}

\definecolor{myblue}{RGB}{0,73,114}
\allsectionsfont{\sffamily\color{myblue}}

\newcounter{myremark}

\newcounter{mynotation}

\usepackage{paralist}

\renewenvironment{itemize}[1]{\begin{compactitem}#1}{\end{compactitem}}
\renewenvironment{enumerate}[1]{\begin{compactenum}#1}{\end{compactenum}}

\newcommand{\mycomment}[1]{}
 \DeclareUnicodeCharacter{202F}{FIX ME!!!!}

\makeatletter
\def\@seccntformat#1{\@ifundefined{#1@cntformat}%
	{\csname the#1\endcsname\quad}  % default
	{\csname #1@cntformat\endcsname}% enable individual control
}
\let\oldappendix\appendix %% save current definition of \appendix
\renewcommand\appendix{%
	\oldappendix
	\newcommand{\section@cntformat}{\appendixname~\thesection\quad}
}
\makeatother

\usepackage{titlesec}

\usepackage{scalerel,stackengine}
\stackMath
\newcommand\reallywidehat[1]{%
\savestack{\tmpbox}{\stretchto{%
  \scaleto{%
    \scalerel*[\widthof{\ensuremath{#1}}]{\kern-.6pt\bigwedge\kern-.6pt}%
    {\rule[-\textheight/2]{1ex}{\textheight}}%WIDTH-LIMITED BIG WEDGE
  }{\textheight}% 
}{0.5ex}}%
\stackon[1pt]{#1}{\tmpbox}%
}

\usepackage[style=apa,
natbib=true,
backend=biber,
maxbibnames=4,
minbibnames=3,   % bibliography: up to 4, then 3 + et al.
maxcitenames=2,  % if total authors <= 2, print them all
mincitenames=1,  % if total authors > 2, print 1 + et al.
url=false,
doi=false,
eprint=false]{biblatex}
\let\cite\textcite

\begin{document}
\setlength{\abovedisplayskip}{0.15cm}
\setlength{\belowdisplayskip}{0.15cm}
\pagestyle{empty}
%\singlespacing
\begin{titlepage}

\title{\bfseries\sffamily\color{myblue}  
	Conjugating Variational Inference for Large Mixed Multinomial Logit Models and Consumer Choice}
\author{Weiben Zhang, Rub\'en Loaiza-Maya, Michael Stanley Smith, Worapree Maneesoonthorn }
\date{\today}
\maketitle
\noindent
{\small Weiben Zhang is a PhD student and Michael Smith is Professor of Management (Econometrics) at the Melbourne Business School, University of Melbourne, Australia. Rub\'en Loaiza-Maya and Worapree Maneesoonthorn are Associate Professors at the Department of Econometrics and Business Statistics, Monash University, Australia. This research was supported by the Commonwealth through an Australian Government Research Training Program Scholarship [DOI: https://doi.org/10.82133/C42F-K220]. Rub\'en Loaiza-Maya, Michael Smith and Worapree Maneesoonthorn gratefully acknowledge support by the Australian Research Council through grants DP200101414, DE230100029 and DP250101069, respectively.  Correspondence should be directed to Michael Smith at {\tt mike.smith@mbs.edu}.
%\noindent \textbf{Acknowledgments:} Here.
}

%\normalsize
\newpage
\begin{center}
\mbox{}\vspace{2cm}\\
{\LARGE \title{\bfseries\sffamily\color{myblue} Conjugating Variational Inference for Large Mixed Multinomial Logit Models and Consumer Choice}}\\
\vspace{1cm}
{\Large Abstract}
\end{center}
\vspace{-1pt}
\onehalfspacing
\noindent
Heterogeneity in multinomial choice data is often accounted for using 
logit models with random coefficients. Such models are called ``mixed'', but they 
can be difficult to estimate for large datasets. 
We review current Bayesian variational inference (VI) methods that can do so, and propose a new VI method that scales more effectively. 
The key innovation is a step that updates efficiently a Gaussian approximation to the conditional posterior of the random coefficients, addressing a bottleneck within the variational optimization. 
The approach is used to estimate three types of mixed logit models: standard, nested and bundle variants. We first demonstrate
the improvement of our new approach over existing VI methods using simulations. Our method is then applied to a large scanner panel dataset of pasta choice. We find consumer response to price and promotion variables exhibits substantial heterogeneity at the grocery store and product levels. Store size, premium and geography are found to be drivers of store level estimates of price elasticities. Extension to bundle choice with pasta sauce improves model accuracy further. 
Predictions from the mixed models are more accurate than those from fixed coefficients equivalents, and our VI method provides insights in circumstances which other methods find challenging.
\vspace{20pt}
 
\noindent
{\bf Keywords}:  Bundles; Consumer Choice; Heterogeneity; Nested Logit; Stochastic Gradient Descent; Random Coefficients; Re-parameterization trick; Variational Bayes.

\end{titlepage}
%\doublespacing

\newpage
\pagestyle{plain}
\setcounter{equation}{0}
\renewcommand{\theequation}{\arabic{equation}}
\renewcommand{\qedsymbol}{\rule{0.7em}{0.7em}}
% !TeX spellcheck = en_US

\section{Introduction}\label{sec:intro}
Heterogeneity is a common feature of multivariate discrete choice datasets, reflecting differences in tastes, preferences, and unobserved factors across observational units; see~\textcite{trainDiscreteChoiceMethods2009} for an overview. Various approaches have been proposed to account for this, including latent class~\citep{kamakura1989probabilistic,greeneLatentClassModel2003} and scale heterogeneity \citep{fiebigGeneralizedMultinomialLogit2010} models. However, random coefficient models (or mixed models) remain one of the most popular approaches due to their 
robustness and interpretation. In particular, the mixed multinomial logit model (MMNL) \citep{mcfaddenMixedMNLModels2000} and its extensions are widely used in
applications in transportation~\citep{bhatQuasirandomMaximumSimulated2001,REN2025103220}, marketing~\citep{danaherAdvertisingEffectivenessMultiple2020,dube2021random}, health economics~\citep{clarkDiscreteChoiceExperiments2014,de2022preferences} and elsewhere. However, Bayesian inference for large MMNL models can be challenging, particularly for high dimensional random coefficient vectors. 

Variational inference (VI) is an approach that has the potential to do so. However, standard VI methods are often not well suited to large MMNL models, and in this paper we first review VI methods that can be used. We then propose a new VI method that we label ``conjugating VI'' (CVI) for estimating large MMNL models, including important variants for nested~\citep{McFadden1978ResidentialLocation}
and bundle~\citep{gentzkowValuingNewGoods2007} choices. Using simulation studies we establish its accuracy against exact posterior inference, computed using Markov chain Monte Carlo (MCMC), for smaller sample sizes. For larger sample sizes of up to 5 million observations with 90,000 random coefficient values, we demonstrate its competitive performance compared to 
the alternative VI methods of~\cite{ongGaussianVariationalApproximation2018b} and \cite{rodriguesScalingBayesianInference2022}.
But the main application of our CVI method is to estimate standard, nested and bundle 
MMNL variants for a large marketing scanner panel dataset. This records over half million sales of pasta from 381 stores of a U.S. grocery chain. Our objective is to capture the store and product-based heterogeneity in the response to price and two promotion variables, and to measure its impact on predictive accuracy. To do so requires estimation of large random coefficient vectors of between 56 and 76 elements. This is a challenging inference problem, particularly when the covariance matrix of the random coefficients is unrestricted as assumed here.

Simulation methods have long been used to estimate mixed logit models~\citep{hajivassiliouMethodSimulatedScores1998, mcfaddenMixedMNLModels2000, bhatQuasirandomMaximumSimulated2001}, including MCMC methods for Bayesian inference \citep{allenbyModelingHouseholdPurchase1994,fruhwirth2007auxiliary, rossiBayesianStatisticsMarketing2003}. However, such approaches can become computationally prohibitive when the number of observations, model size and dimension of the random coefficients are high. 
Developments in P\'olya-Gamma augmentation techniques \citep{polsonBayesianInferenceLogistic2013,zensUltimatePolyaGamma2024} improve the efficiency of MCMC methods, but can still encounter bottlenecks in complex large examples. In addition,
estimation by Hamiltonian Monte Carlo \citep{hoffman2014no} scales poorly with the dimension of the random coefficients vector because it evaluates the gradient of the log posterior with respect to it.

VI methods are an increasingly popular alternative to MCMC for computing Bayesian inference for large statistical models; see~\cite{loaiza-mayaFastAccurateVariational2022}, \cite{chan2022fast}, \cite{BernardiBianchiBianco2024}, \cite{PrueserHuber2024} and~\cite{KorobilisSchroeder2025} for some recent econometric applications. 
VI approximates the posterior using a family of tractable distributions. The variational approximation (VA) is the member of the family that minimizes the Kullback Leibler divergence (KLD) between it and the exact posterior. 
Effective VI methods balance speed and accuracy, the key to which is selection of the approximating family and optimization method. The most common VAs use mean-field approximations that assume independence across partitions of model parameters, which can be restrictive, but the resulting optimization can be solved using very fast coordinate-ascent algorithms. \cite{goplerudFastAccurateEstimation2022,goplerudPartiallyFactorizedVariational2025} propose efficient coordinate–ascent VI for standard MMNL models using P\'olya–Gamma augmentation. However, 
their efficiency hinges on variational families and priors that admit closed-form updates, and application to some MMNL specifications, such as nested logit models, are infeasible within this framework.

More flexible variational families can be used by adopting efficient stochastic gradient optimization algorithms~\citep{hoffmanStochasticVariationalInference2013}. One effective strategy for MMNL models is data augmentation VI (DAVI) which applies such an approach to the posterior augmented with the random coefficients; for example, see \cite{tanVariationalInferenceGeneralized2013}, \cite{ranganathBlackBoxVariational2014} and~\cite{ongGaussianVariationalApproximation2018b}. However, this approach may not capture the dependence between the random coefficients and model parameters accurately; see the discussion in~\cite{loaiza-mayaFastAccurateVariational2022}. Another alternative is to use amortized VI (AVI), which has been applied to MMNL models by~\cite{rodriguesScalingBayesianInference2022}. AVI uses a neural network as an auxiliary model to approximate the posterior of the random effects. In this paper we compare the efficacy of these two current VI approaches when estimating the standard, nested and bundle variants of the MMNL for large scale problems. 

We also propose a new approach tailored to the estimation of MMNL models. Our VA employs a second-order Taylor expansion of the likelihood which, combined with a Gaussian hierarchical prior, yields a Gaussian approximation to the random coefficients posterior. This conjugacy is why we label the method CVI, and show it can be a more accurate VA than DAVI. In related work, \cite{tanUseModelReparametrization2021} also employs a second-order Taylor expansion, but our method differs from theirs because we treat the expansion centers as auxiliary variational parameters that are updated infrequently and without optimization. This avoids the need to solve a numerical optimization problem for each random coefficient group at every iteration as in \cite{tanUseModelReparametrization2021}, so that CVI is more scalable.

CVI is used to fit the mixed logit models to household level purchases of 
pasta in this study. The models capture extensive heterogeneity at the store and product level for price and promotion variables. They have greatly improved fit and predictive accuracy compared to equivalent models with fixed coefficients. We show how to segment the variation by source, and find that it is mostly due to branding rather than pasta type, which is consistent with prior marketing studies; e.g. see~\cite{ChinJMR1991}. Using estimates of store level price elasticities, we also find these are strongly related to the store size, location and positioning, with customers of larger stores that are value-positioned exhibiting greatest (i.e. more negative) price elasticity. A major contribution of the paper is to use CVI to estimate a model that also allows for bundling of pasta purchases with those of pasta sauce. Bundle MMNLs are more difficult to estimate because the sample, parameters and random coefficients are typically of greater size. We find that including extra information on pasta sauce improves predictive accuracy for purchases of pasta. But this is only observed in the mixed case, and is not observed for fixed coefficients. 

The rest of the paper is organized as follows. Section~\ref{sec: mixedlogit} outlines the three different MMNL specifications, along with suitable priors. A brief introduction to VI follows, with a focus on those approaches suitable to estimate MMNL models. Section~\ref{sec:cvi} presents the proposed CVI method. Section~\ref{sec:sim} contains the simulation studies and Section~\ref{sec: empirical} the consumer choice application; Section~\ref{sec: disc} concludes.

\section{Variational Inference for Mixed Multinomial Logit Models}\label{sec: mixedlogit}
In this section, we first specify three mixed multinomial logit models for which VI methods will be used. 
It is common to assume the random coefficients are normally distributed, and we focus on this case here. Parameter identification and priors are then discussed, followed
by a brief overview of VI, including two recently suggested VI methods that can be used to estimate these choice models when the random coefficients are of high dimension. 

\subsection{Three mixed multinomial logit models}

\paragraph{MMNL:} The first model is the MNL~\citep{mcfaddenConditionalLogitAnalysis1974}
with random coefficients to capture unobserved preference heterogeneity across groups~\citep{mcfaddenMixedMNLModels2000}.
Consider a group $i\in \{1, \dots, S\}$ with choice set $C_{it}\subseteq\{1,2,\ldots,J\}$ on choice occasion $t \in \{1, \dots, T_i\}$. The choice set $C_{it}$ may vary across groups and occasions. Define the utility of alternative $j \in C_{it}$ to be $u_{itj} = v_{itj} + \varepsilon_{itj}$, 
where $\varepsilon_{itj}$ is i.i.d. Gumbel distributed. Here, $v_{itj}$ consists of both fixed and random components~\footnote{The observed component of the total utility is also referred to as the ``representative utility''. For simplicity, we will refer to  $v_{itj}$  as the utility throughout the rest of the paper.}:
\begin{align}
	v_{itj} = \betavec^\top \xvec^f_{itj} + {\alphavec_{ij}^\top} \xvec^r_{itj} \label{v_itj_mmnl},
\end{align}
where $\xvec^f_{itj}$ denotes an $(w^f\times 1)$ covariate vector with fixed coefficients $\betavec$, and $\xvec^r_{itj}$ denotes an $(w^r\times 1)$ covariate vector with group-specific random coefficients {$\alphavec_i  = (\alphavec_{i2}^\top, \dots, \alphavec_{iJ}^\top)^\top \sim N(\xivec,\Sigma)$}. For identification purposes $\alphavec_{i1}=\zerovec$ as discussed further below. 
% 
%{The mean $\xivec$ is a vector of length $w = w^rJ$ and $\Sigma$ is a $w \times w$ full covariance matrix that captures the correlation among random coefficients across alternatives.}
If
%$\xvec_{itj} = \left((\xvec^f_{itj})^\top ,(\xvec^r_{itj})^\top\right)^\top$, and 
$Y_{it}$ is the choice outcome of group $i$ at occasion $t$, and $\xvec_{itj} = \left((\xvec^f_{itj})^\top ,(\xvec^r_{itj})^\top\right)^\top$ are the covariate values, then it is straightforward to show that
\begin{align}
	\mbox{Pr}(Y_{it} = j\mid \xvec_{it1},\ldots,\xvec_{itJ}) = \frac{\exp(v_{itj})}{\sum_{j \in C_{it}} \exp(v_{itj})}\,,\;\mbox{ for }j\in C_{it}\,. \label{p_mmnl}
\end{align}
 The model parameters $\thetavec = (\betavec^\top, \xivec^\top,\mbox{vech}(\Sigma)^\top)^\top$ and latent variables $\alphavec = (\alphavec_1^\top, \dots, \alphavec_S^\top)^\top$, where $\mbox{vech}(\Sigma)$ is the half-vectorization of $\Sigma$. In the machine learning literature it is common to call $\thetavec$ global parameters, and $\alphavec$ local parameters~\citep{hoffmanStochasticVariationalInference2013} as we sometimes do here.

\paragraph{B-MMNL:} There is a growing interest in modeling the roles of bundles and 
complements in microeconomics~\citep{gentzkowValuingNewGoods2007,sun2024bundle}, transportation~\citep{caiatietal2020} and marketing~\citep{chungGeneralChoiceModel2003}. 
Our second model is a variant of that proposed by~\cite{gentzkowValuingNewGoods2007} that  
generalizes the MMNL above to also allow for bundles, and we label it ``B-MMNL''. Consider the case where there are $R$ choices including bundles. For example, if there were $J=3$ alternatives
A,B and C, which could be selected singularly or also as bundles \{A,B\}, \{B,C\}, \{A,C\} and \{A,B,C\}, then $R=7$. At each choice opportunity, the choice set $C_{it}\subseteq \{1,\ldots,R\}$ is expanded to include any bundles, and can vary over group $i$ and occasion $t$. For $r=1,2,\ldots,R$, denote the members of the bundle as ${\cal B}_r\subseteq\{1,\ldots,J\}$.\footnote{Continuing the example where the $R=7$ choices are 
	A, B, C, \{A,B\}, \{B,C\}, \{A,C\} and \{A,B,C\} labeled as $1,2,\ldots,7$ in this order.   
If choice $r=1$ corresponds to the singleton \{A\}, then ${\cal B}_1=1$, whereas if choice $r=5$ corresponds to \{B,C\}, then ${\cal B}_5=\{2,3\}$.}
The utility of alternative $r\in C_{it}$ is
\begin{align}
	v_{itr} = \sum_{j \in {\cal B}_r} \left(\betavec^\top \xvec^f_{itj} + {\alphavec_{ij}^\top} \xvec^r_{itj}\right) + \gamma_r\,, \label{v_itj_mixbc}
\end{align}
where $\gamma_r$ is a complementary effect that is equal to zero if and only if alternative $r$ is a singleton (i.e. not a bundle). With this utility, the choice probability is
\begin{align}
	\mbox{Pr}(Y_{it}=r\mid \xvec_{it1},\ldots,\xvec_{itJ}) = \frac{\exp(v_{itr})}{\sum_{l\in C_{it}}\exp(v_{itl})}\,,\;\mbox{ for }r\in C_{it}\,. \label{p_mixbc}
\end{align} 
The model parameters $\thetavec$ also include the complementary effects $\gamma_r$ for all alternatives $r$ that correspond to bundles and not singletons.

\paragraph{MNestL:}
The third model is a mixed variant of the classical nested logit (MNestL) model~\citep{McFadden1978ResidentialLocation}. This model accounts for a sequential choice process, which behavioral research suggests can better reflect decision-making processes~\citep{kovach2022behavioral}. The choice set is partitioned into $K$ disjoint nests $C_{it}=\bigcup_{k=1}^K B_{itk}$, and the decision maker first selects a nest, followed by an
alternative within the nest. Marginalizing over the sequential decision, the
utility function is specified as at~\eqref{v_itj_mmnl}, but where
$\varepsilon_{itj}$ is distributed generalized extreme value (GEV). 
The probability of each alternative $j\in  B_{itk}$ and nest $k=1,\ldots,K$ is
\begin{align}
	\mbox{Pr}(Y_{it}=j\mid \xvec_{it1},\ldots,\xvec_{itJ})= \frac{\exp (v_{itj}/\tau_k) \left(\sum_{a \in B_{itk}}\exp(v_{ita}/\tau_k)\right)^{\tau_k -1}}{\sum_{l = 1}^{K}\left(\sum_{b \in B_{itl}}\exp(v_{itb}/\tau_l)\right)^{\tau_l }}\,,
	\label{p_mixnl}
\end{align}
with 
nesting parameters $\tauvec = (\tau_1, \dots, \tau_K)^\top$ that measure the degree of dependence in each of the $K$ nests;
see~\citet[Sec. 4.1]{trainDiscreteChoiceMethods2009} for an introduction to this model. 
The parameters $\thetavec$
now include $\tauvec$, and 
when $\tauvec = (1,1,\ldots,1)$ the MNestL model degenerates to 
the MMNL model.

\paragraph{Identification:} To identify all three choice models, we take alternative $j=1$ as a reference, and normalize utilities via the differences $\tilde{u}_{itj} = u_{itj} - u_{it1}$ for all $j$. This is equivalent to subtracting $\xvec_{it1}$ from the covariates of each alternative, and setting the coefficients of the reference alternative to zero; see~\citet[Sec.~2.5]{trainDiscreteChoiceMethods2009}. The dimension of $\alphavec_i$ is therefore $w = w^r(J-1)$, because the random coefficients associated with $\xvec^r_{it1}$ are fixed to zero. {Therefore,  $\xivec$ is a vector of length $w$, and $\Sigma = \{\Sigma_{j\ell}\}$ is a $w \times w$ full covariance matrix, where each block  $\Sigma_{j\ell}=\mbox{Cov}(\alphavec_{j+1},\alphavec_{\ell+1})$ is of size $w^r \times w^r$.} 
Many studies fix $\Sigma_{j\ell}=\mathbf{0}$ for $j \neq \ell$, or restrict $\Sigma$ to be diagonal, whereas we instead estimate an unconstrained $\Sigma$ matrix, which is much more challenging. The likelihoods of the three models are the products of their respective mass functions evaluated at the observed data.

\paragraph{Priors:} We assign weakly informative Gaussian priors $N(0, 100)$ to each element of $\betavec$, $\xivec$, and $\gamma_r$. For the elements of 
$\tauvec$, we use independent half-t priors with scale 1.5 and degrees of freedom 5. This prior places substantial mass in the interval [0, 1] while maintaining heavy tails. But in a mixed model, the prior on 
%\sout{ $\Sigma=\{\sigma_{ij}\}$ for $i\neq j$} 
$\Sigma$ is particularly impactful, and there are two popular choices. The first was proposed by~\cite{lewandowskiGeneratingRandomCorrelation2009} and is for the decomposition $\Sigma=T\Omega T$, where $\Omega$ is the correlation matrix and $T=\mbox{diag}(\sigma_{11},\ldots,\sigma_{w w})$ is a diagonal matrix of standard deviations. Then $\sigma_{ll}\sim \mbox{Half-Cauchy}(0,10)$ and a uniform prior $p(\Omega)\propto \mbox{constant}$ is used for the correlation matrix. The second prior was suggested by~\cite{huangSimpleMarginallyNoninformative2013a}, which has density
	\begin{align*}
	p(\Sigma) \propto 
	|\Sigma|^{-(\nu + 2w)/2}
	\prod_{l=1}^{w}
	\left\{
	\nu (\Sigma^{-1})_{ll} + \frac{1}{A_l^2}
	\right\}^{-(\nu + w)/2}.
\end{align*} 
	They show that $\nu = 2$ leads to a marginally uniform distribution over the correlation terms, while large values of $A_l$ leads to weakly informative priors. Here we use $\nu = 2$ and $A_l = 100$ for all $l = 1,\dots, w$. We refer to the two priors as ``LKJ'' and ``HW'', with the LKJ prior uniform over the space of correlation matrices, and the HW prior marginally uniform over each correlation coefficient. Finally, when there are a very large number of alternatives (as in some bundle models) a prior based on a factor model for $\Sigma$ can also be adopted~\citep{murray2013bayesian}, so that the number of parameters only increases linearly.

\subsection{Variational inference}
Let $\bm{y}$ denote the observed choice data, $p(\yvec|\psivec)$ be the likelihood
and $p(\psivec)$ the prior, 
%and set $\psivec=(\thetavec^\top,\zvec^\top)^\top$ to the model parameters augmented by any 
%latent variables.
then
VI approximates the Bayesian posterior density $p(\psivec| \yvec) \propto p(\yvec |\psivec)p(\psivec)\equiv g(\psivec)$ with a simpler density $q_\lambda(\psivec)\in {\cal Q}$ called the variational approximation (VA), where ${\cal Q}$ is a family of flexible but tractable densities. 
Here, $\lambdavec$ are called ``variational parameters'' that fully characterize the VA,
and their values are obtained by minimizing the divergence measure between $q_\lambda(\psivec)$ and $p(\psivec|\yvec)$. The KLD is the most popular choice, and it is easily shown that minimizing the KLD corresponds to maximizing the evidence lower bound (ELBO):  
\begin{align*}
	\calL = E_{q_\lambda}\left[\log g(\psivec) - \log q_\lambda(\psivec)\right] 
	%\label{elbo}
\end{align*}
with respect to $\lambdavec$. The expectation above is with respect to $\psivec\sim q_\lambda(\psivec)$.

For large scale models, it is popular to solve the optimization problem using stochastic gradient descent (SGD)~\citep{bottouLargeScaleMachineLearning2010}
which requires an unbiased approximation to the gradient  $\nabla_{\lambda}\calL(\lambdavec)$. This can be computed efficiently using the re-parameterization trick \citep{rezendeStochasticBackpropagationApproximate2014,kingmaAutoEncodingVariationalBayes2014}. It is based on the re-parameterization $\psivec = h(\epsilonvec,\lambdavec)$, where $h$ is a deterministic function and $\epsilonvec \sim f_{\epsilon}$ with distribution $f_{\epsilon}$ is invariant to $\lambdavec$. Then the gradient is
\begin{align}
	\nabla_{\lambda}\calL(\lambdavec) = E_{f_\epsilon}\left[\frac{\partial h(\epsilon,\lambdavec)^\top}{\partial \lambdavec}\nabla_{\psivec}\calL(\psivec)\right], \label{equ:reparameterized_gradient}
\end{align}
and the expectation can be approximated using one or more Monte Carlo draws from $f_\epsilon$. %In this paper, for the Gaussian approximations we adopt, we set $\epsilonvec\sim~N(\bm{0},I)$.
For recent overviews of VI see~\cite{bleiVariationalInferenceReview2017b} and~\cite{tran2021practical}. 

When estimating large mixed logit models that are the subject of this paper, 
the target posterior is augmented with the latent random coefficients, so that 
$\psivec=(\thetavec^\top,\alphavec^\top)^\top$, $p(\yvec|\psivec)$ is an extended likelihood and 
the prior $p(\psivec)=p(\alphavec|\thetavec)p(\thetavec)$. 
For large models---particularly when 
the dimension of each latent variable vector {$\alphavec_i$} is high---direct application of generic VI methods is difficult. Below we discuss two approaches that are specifically 
tailored for this case.

\mycomment{
The stochastic VI is more flexible with the choice of priors and often does not need factorization for the model parameters. Stochastic VI often involves a partially factorized approximation for mixed models, where the random effects are independent across groups:
\begin{align}
	q(\psivec) = q_\lambda(\thetavec)\prod_{i  = 1}^{S}q(\zvec_i\mid \muvec_i, \Sigma_i)
	\label{MFVI}
\end{align}
where $\lambdavec$ is the variational parameters for the model (global) parameters and $\muvec_i$ and $\Sigma_i$ governs the distribution of random effects of group $i$. We now outline two recently developed VI methods that are applicable to mixed logit models: the data augmentation VI (DAVI) and amortized VI (AVI). 
}

\paragraph{DAVI:} Following~\cite{hoffmanStochasticVariationalInference2013}, it is common to use a factorized mean-field VA for latent variable models, where $\thetavec$ and each latent variable realization are independent. We label such a method as data augmentation VI (DAVI), and the approach has been widely used
when computing VI for random coefficient models, including by~\cite{tanVariationalInferenceGeneralized2013}, \cite{menictasStreamlinedVariationalInference2023},  \cite{danaherAdvertisingEffectivenessMultiple2020} and \cite{goplerudPartiallyFactorizedVariational2025} among others. For the mixed multinomial logit models in the current paper, the VA is
\begin{align}
	q_\lambda(\psivec) = q_{\lambda_0}(\thetavec)\prod_{i  = 1}^{S}q_{\lambda_i}(\alphavec_i)\,.
	\label{MFVI}
\end{align}
In our implementation we set $q_{\lambda_0}(\thetavec) = \phi(\thetavec;\muvec_0,V_0)$ and $q_{\lambda_i}(\alphavec_i) = \phi(\alphavec_i;\muvec_i,V_i)$ for $i \in \{1, \dots, S\}$, where $\phi(\xvec;\muvec_i,V_i)$ is a Gaussian density with mean $\muvec_i$ and covariance $V_i$ evaluated at $\xvec$.
The variational parameters are $\lambdavec=(\lambdavec_0^\top,\ldots,\lambdavec_S^\top)$ with  $\lambdavec_i=(\muvec_i^\top,\mbox{vech}(V_i)^\top)^\top$. 

The matrix $V_i$ is $(w \times w)$ for $i\geq 1$ and $V_0$ is $(w_0\times w_0)$, so that 
the number of variational parameters increases quadratically with the dimension of $\alphavec_i$ and/or $\thetavec$. We follow~\cite{ongGaussianVariationalApproximation2018b} and use a factor model  (also called a ``low-rank plus diagonal'') for these matrices\footnote{The approximate posterior covariance matrices of the random coefficients, $V_1,\ldots,V_S$, should not be confused with the prior covariance matrix of the random coefficients, $\Sigma$, which are all of the same size {$(w \times w)$.}} with
\begin{align}\label{Eq:FACTOR}
	V_i = B_iB_i^\top  + \text{diag}(\dvec_i^2) \quad \text{ for }  i \in \{0, \dots, S\}\,,
\end{align}
where is $B_i$ a loadings matrix with $p$ columns, 
and $\dvec_i$ is a vector. Following these authors, we set the upper triangle of $B_i$ to 
zero and $p=5$ in our empirical work, which these authors show is a good choice in a range of examples.\footnote{Unlike when using a factor model for modeling observed data, it is not necessary nor efficient for $B_i,\dvec_i$ to be uniquely identified; see the discussion in~\cite{ongGaussianVariationalApproximation2018b}.} 
The variational parameters are now $\lambdavec_i=(\muvec_i^\top, \text{vech}(B_i)^\top,\dvec_i^\top)^\top$ for $i\in \{0, \dots, S\}$, which is much lower dimension than when $V_0,\ldots,V_S$ are unrestricted.
The derivatives in~\eqref{equ:reparameterized_gradient} for implementation of efficient SGD are given in \cite{ongGaussianVariationalApproximation2018b}. 
While this approach is fast and scalable, the assumption of independence between $\thetavec$ and $\alphavec$ is often unrealistic and can lower accuracy;
see Part~B.1 of the Online Appendix for the DAVI algorithm.

\paragraph{AVI:} 
Amortized variational inference (AVI)  \citep{zhangAdvancesVariationalInference2019b,margossianAmortizedVariationalInference2024} is an emerging approach that aims to avoid the proliferation of variational parameters across heterogeneous units, such as the random coefficients here. 
AVI trains a parametric function---usually a neural network (NN)---with parameters $\Psi$ that maps the local data ${\cal D}_i=\{(y_{it},\xvec_{itj});t=1,\ldots,T_i,\,j=1,\ldots,J\}$ to $\lambdavec_i$ for $i=1,\ldots,S$. The parameters $\Psi$ and $\lambdavec_0$ (the parameters of the VA of $\thetavec$) are learned jointly by maximizing the ELBO using SGD with the re-parameterization trick as discussed previously. 
Because the same NN is used for all groups, the number of variational parameters that require learning does not increase with $S$, allowing application to datasets with large numbers of groups. 

\cite{rodriguesScalingBayesianInference2022} uses AVI to estimate the MMNL model, where
 an NN with a convolution layer, max pooling, batch normalization and one fully connected layer is used to predict $\muvec_i$ and $V_i$ of the Gaussian approximation to each $\alphavec_i$. The VA is factorized as 
\[
	q_\lambda(\psivec) = q_{\lambda_0}(\thetavec) \prod_{i = 1}^{S}\phi\left(\alphavec_i;\muvec_i({\cal D}_i,\Psi),V_i({\cal D}_i,\Psi)\right).
\]
where we write $\muvec_i,V_i$ as functions of $\Psi$ and ${\cal D}_i$ to make clear that they depend on the NN auxiliary model.
The author shows that the accuracy of AVI
is competitive compared to a maximum simulated likelihood estimator (MSLE) \citep{bhatQuasirandomMaximumSimulated2001} and an MCMC method, and that AVI is more scalable than the two other methods.

\section{Conjugating variational inference}\label{sec:cvi}
We now outline a new VI approach as an alternative estimation strategy for mixed effects multinomial logistic models. To approximate the posterior distribution  we use the variational family $\mathcal{Q} = \{q_{\lambda_0}(\bm{\psi}| \bm{\vartheta},\bm{a}):\bm{\lambda}_0\in \Lambda_0\}$ with elements defined as
\begin{align}\label{Eq:approx}
	q_{\lambda_0}(\bm{\psi}| \bm{\vartheta},\bm{a})
	= q_{\lambda_0}(\bm{\theta})
	\prod_{i=1}^{S}
	q\big(\bm{\alpha}_i | \bm{\vartheta},\bm{a}_i \big).
\end{align}
For $q_{\lambda_0}(\bm{\theta})$ we use the same Gaussian variational approximation with a factor covariance structure as with DAVI at \eqref{Eq:FACTOR} so that  $\lambdavec_0=(\muvec_0^\top, \text{vech}(B_0)^\top,\dvec_0^\top)^\top$.
Our VA of the random coefficients introduces two sets of auxiliary parameters: $\bm{a}^\top = (\bm{a}_1^\top,\dots,\bm{a}_S^\top)\in \mathbb{R}^{Sw}$ and $\bm{\vartheta} \in \Theta$ is in the same parameters space as the global parameters $\thetavec$.  As demonstrated in \cite{loaiza-mayaFastAccurateVariational2022}, for a given choice of approximation $q_{\lambda_0}(\bm{\theta})$ for the global parameters, the best approximation for the random coefficients is the conditional posterior $p(\bm{\alpha}_i|\bm{y}_i,\bm{\theta})\propto 
p(\bm{y}_i | \bm{\theta},\bm{\alpha}_i)p(\bm{\alpha}_i | \bm{\theta}),$ which yields a variational error equal to that obtained when integrating out $\bm{\alpha}_i$ exactly.  However, in multinomial logistic models this conditional is unavailable in closed form and cannot be generated from efficiently. 
Guided by this observation, we choose $q\big(\bm{\alpha}_i | \bm{\vartheta},\bm{a}_i \big)$ to closely mimic 
$p(\bm{\alpha}_i| \bm{y}_i,\bm{\theta})$ by introducing ``approximate conjugacy'', which we construct as follows. We construct an approximation to the random coefficients conditional on the proxy parameter $\bm{\vartheta}$ (instead of $\thetavec$), which  will reduce the computational cost of calibrating the approximation.
This replaces the log-likelihood $\log p(\bm{y}_i | \bm{\vartheta},\bm{\alpha}_i)$ conditioning on $\bm{\vartheta}$ with its
second-order Taylor expansion around $\bm{a}_i$,
\[
\log \tilde{p}(\bm{y}_i | \bm{\vartheta},\bm{\alpha}_i)
=
\log p(\bm{y}_i | \bm{\vartheta},\bm{a}_i)
+ \bm{g}_i^\top(\bm{\alpha}_i-\bm{a}_i)
- \tfrac{1}{2}(\bm{\alpha}_i-\bm{a}_i)^\top H_i (\bm{\alpha}_i-\bm{a}_i),
\]
where $\bm{g}_i$ and $H_i$ are the gradient and (negative) Hessian of 
$\log p(\bm{y}_i | \bm{\theta},\bm{\alpha}_i)$ evaluated at $\bm{\alpha}_i=\bm{a}_i$ and $\bm{\theta}=\bm{\vartheta}$.
The conjugate approximation is then
\begin{equation}\label{Eq:CVIREapprox}
q\big(\bm{\alpha}_i | \bm{\vartheta},\bm{a}_i \big)\;\propto\; 
\tilde{p}(\bm{y}_i | \bm{\vartheta},\bm{\alpha}_i)\, p(\bm{\alpha}_i | \bm{\vartheta}),
\end{equation}
which is a multivariate normal distribution so that $q\big(\bm{\alpha}_i | \bm{\vartheta},\bm{a}_i \big) = \phi(\alphavec_i; \muvec_i, V_i),$
with $V_i= (H_i+ \Sigma^{-1})^{-1}$ and $\muvec_i= V_i (\vvec_i + \Sigma^{-1}\xivec)$ and $\vvec_i = \bm{g}_i + H_i \avec_i$;  see part~B.3 of the Online Appendix. Note that the quantities $\muvec_i$, $V_i$, $\bm{g}_i$ and $H_i$ are all functions of $\bm{y}_i$, $\bm{a}_i$ and $\bm{\vartheta}$.

Consider the Gaussian generative formula from $q_{\lambda_0}(\bm{\theta})$ such that $\bm{\theta}=h(\bm{\epsilon},\bm{\lambda})= \bm{\mu}_0+B_0\bm{z}+\bm{d}_0\circ\bm{e}$ with $\bm{\epsilon} = (\bm{e}^\top,\bm{z}^\top)^\top$, $\bm{e}\sim N(\bm{0}_{w_0},I_{w_0})$ and $\bm{z}\sim N(\bm{0}_p,I_p)$. Denote $f_{e,z}(\bm{e},\bm{z})$ to be the multivariate normal density function of $\bm{e}$ and $\bm{z}$. Using the re-parametrization trick of \cite{kingmaAutoEncodingVariationalBayes2014}, it is possible to show that the ELBO gradient for the VA in \eqref{Eq:approx} can be written as
\begin{equation}\label{Eq:ELBO}
\nabla_{\lambda_0}\mathcal{L}\left(\bm{\lambda}_0\right)  = E_{f_{e,z}(e,z)q(\bm{\alpha}|\bm{\vartheta},\bm{a})}\left[\frac{\partial\bm{\theta}}{\partial\bm{\lambda}_0}^\top\left\{\nabla_\theta\log g(\bm{\psi})-\nabla_\theta\log q_{\lambda_0}(\bm{\theta})\right\}\right].
\end{equation}
See Online Appendix B.4 for a derivation of this gradient. An unbiased single-sample estimator of this gradient is obtained by drawing once from $\bm{e}\sim N(\bm{0}_{w_0},I_{w_0})$, $\bm{z}\sim N(\bm{0}_p,I_p)$ and $\bm{\alpha} \sim q(\bm{\alpha}| \bm{\vartheta},\bm{a})$,  and evaluating the ELBO gradient at that draw; we denote this estimator by 
$\widehat{\nabla_{\lambda_0}\mathcal{L}(\bm{\lambda}_0)}$.  Computing it requires only the gradient of the augmented posterior and the term  $\nabla_{\bm{\theta}}\log q_{\lambda_0}(\bm{\theta})$, which is available in closed form (see, e.g., \citealt{ongGaussianVariationalApproximation2018b}).  

Crucially, by conditioning the random coefficients approximation on the proxy parameter $\bm{\vartheta}$ rather than on the full parameter vector $\thetavec$, the gradient no longer contains terms of the form $\nabla_{\bm{\theta}} \log q(\bm{\alpha}_i | \bm{\theta}, \bm{a}_i)$ and therefore avoids derivatives such as $\frac{\partial^3}{\partial\bm{\alpha}_i\partial\bm{\alpha_i}\partial \bm{\theta}} \log p(\bm{y}_i | \bm{\vartheta},\bm{\alpha}_i)  $, which are   expensive to compute. In addition, conditioning on $\bm{\vartheta}$ removes the need to evaluate $\bm{g}_i$ and $H_i$ at every iteration of the calibration approach described below, leading to a substantial computational saving.
This design choice is in fact a key factor that distinguishes our method from that of \cite{tanUseModelReparametrization2021}. Their approach also relies on Taylor expansions to approximate the random coefficients posterior, but each iteration of their VI algorithm requires a nested optimization to determine around which random coefficients values to center the Taylor expansions. Moreover, because their approximation explicitly depends on $\thetavec$, the gradients involve computationally costly terms related to $H_i$, further increasing computational complexity. As a consequence, their method becomes significantly more burdensome when either the dimension of $\alphavec_i$ or the number of groups $S$ is large.

The variational parameter vector $\bm{\lambda}_0$ is calibrated via SGA by iterating over 
\begin{equation}\label{SGA}
	{\bm{\lambda}_0}^{(j+1)} = {\bm{\lambda}_0}^{(j)}+\bm{\rho}^{(j)}\circ \widehat{\nabla_{\lambda_0}\mathcal{L}(\bm{\lambda}_0^{(j)})}
\end{equation}
until reaching convergence. The symbol $\circ$ denotes the Hadamard (element-wise) product, and the vector $\bm{\rho}^{(j)}$ represents the adaptive step size, which we set following the ADADELTA method of \cite{zeilerADADELTAAdaptiveLearning2012a}. The proxy auxiliary parameter vector is updated every $\kappa$ iterations to match the current variational mean, $\bm{\vartheta}^{(j+1)} = \bm{\mu}^{\left(\kappa\left\lfloor \frac{j+1}{\kappa} \right\rfloor\right)},$
which helps ensure that $\bm{\vartheta}$ remains a good proxy for typical draws from $q_{\lambda_0}(\thetavec)$. Each update of $\bm{\vartheta}$ requires recomputing the quantities $\muvec_i$, $V_i$, $\bm{g}_i$, and $H_i$. Hence, a small $\kappa$ increases the computational cost per iteration but may improve convergence; conversely, a large $\kappa$ reduces per-iteration cost but may slow convergence. In practice, we start  $\kappa=20$ and update it every 500 iterations to be $\kappa = \left\lfloor 1.1\kappa \right\rceil$, where $\left\lfloor\cdot\right\rceil$ denotes the closest integer notation. We found that this choice yields frequent updates in the early stages of the algorithm---when rapid convergence is most critical---and increasingly infrequent updates later on, when the variational mean has largely stabilized and computational efficiency becomes more important. This choice provides a good balance between convergence speed and computational cost. 

\begin{algorithm}[b!]
	\begin{algorithmic}
		\State Initiate ${\lambdavec_0}^{(0)}$, $\avec^{(0)}$, $\varthetavec^{(0)}$, compute $\{\muvec_i^{(0)}\}_{i=1}^S$, $\{V_i^{(0)}\}_{i=1}^S$ and set $j = 0$. 
		\Repeat\\
		\State (a) Generate $\bm{e}^{(j)} \sim N(\bm{0}_{w_0},I_{w_0})$, $\bm{z}^{(j)}\sim N(\bm{0}_{p},I_{p})$  and set $\thetavec^{(j)} =\bm{\mu}_0^{(j)}+B_0^{(j)}\bm{z}^{(j)}+\bm{d}_0^{(j)}\circ\bm{e}^{(j)}$.\\
		
		\State (b) For $i = 1, \dots, S$, generate ${\alphavec_i}^{(j)} \sim  N(\muvec_i^{(j)},V_i^{(j)})$.\\
		
		\State (c) Compute $\nabla_{\lambda_0}\widehat{\mathcal{L}({\lambdavec_0}^{(j)})} = \frac{\partial \thetavec}{\partial \lambdavec_0}^\top \bigl\lvert_{{\lambda_0} = {\lambda_0}^{(j)}} \times \left[\nabla_\theta \log g(\psivec^{(j)}) - \nabla_\theta \log q_{\lambda_0} (\thetavec^{(j)})\right]$.\\
		
		\State (d) Compute step size $\boldsymbol{\rho}^{(j)}$ using an adaptive method (e.g. an ADA method).\\
		
		\State (e) Set ${\lambdavec_0}^{(j+1)} = {\lambdavec_0}^{(j)} + \boldsymbol{\rho}^{(j)} \circ \nabla_{\lambda_0}\widehat{\mathcal{L}({\lambdavec_0}^{(j)})}.$\\
		\ \ \ \ \ \ \ \ \ Set $\bm{\vartheta}^{(j+1)} = \bm{\mu}^{\left(\kappa\left\lfloor \frac{j+1}{\kappa} \right\rfloor\right)}.$\\
		\ \ \ \ \ \ \ \ \ Set $\bm{a}^{(j+1)}= \bm{a}^{(j)}+ r \bigl(\bm{m}^{(j)} - \bm{a}^{(j)}\bigr)\, \mathbb{I}\{\, j+1 = 0 \pmod{\kappa} \,\}$.\\
		
		\State (f) If $\bm{\vartheta}^{(j+1)}$ and $\bm{a}^{(j+1)}$ in step (e) have changed, update $\{\muvec_i^{(j+1)}\}_{i=1}^S$, $\{V_i^{(j+1)}\}_{i=1}^S$.\\
		
		\State (g) Set $j = j+1$.\\
		
		\Until{either a stopping rule is satisfied or a fixed number of steps is taken}
	\end{algorithmic}
	\caption{Conjugating variational inference}
	\label{alg:cvi}
\end{algorithm}

The auxiliary parameter vector $\bm{a} = (\bm{a}_i^\top,\dots,\bm{a}_S^\top)^\top$ is also updated every $\kappa$ iterations as
\[
\bm{a}^{(j+1)}
= \bm{a}^{(j)}
+ r \bigl(\bm{m}^{(j)} - \bm{a}^{(j)}\bigr)
\, \mathds{1}\{\, j+1 = 0 \pmod{\kappa} \,\},
\]
where $\bm{m} = (\bm{m}_1^\top, \dots, \bm{m}_S^\top)^\top$ and $\mathds{1}\{\, j+1 = 0 \pmod{\kappa} \,\}$ equals one if $j+1$ is a multiple of $\kappa$, and zero otherwise.
This smoothed update gradually aligns the Taylor expansion centers with the variational mean of the random coefficients. Once the algorithm has converged(so that $\bm{a}^{(j+1)} = \bm{a}^{(j)}$) we obtain $\bm{a}^{(j+1)} \approx \bm{m}^{(j+1)}$, as desired. To avoid numerical issues when initializing $\avec^{(0)}$, we perform a short 20-iteration warm-up: starting from $\tilde{\avec}^{(0)} = \mathbf{0}$, we iterate $\tilde{\avec}^{(k)} = (1 - r)\tilde{\avec}^{(k-1)} + r\,\bm{m}^{(k-1)}$ for $k = 1,\dots,20$, and then set $\avec^{(0)} = \tilde{\avec}^{(20)}$. We use $r = 0.1$ throughout all examples. Algorithm \ref{alg:cvi} provides a summary on the implementation of our method. To stop the algorithm we use the rule proposed in \cite{rodriguesScalingBayesianInference2022}, which is given in Algorithm \ref{alg:stop_rule} of the Online Appendix.

\section{Simulation Study}\label{sec:sim}
We undertake a simulation study to demonstrate that the VI methods provide fast and accurate estimates 
of the three mixed multinomial logit models outlined. We establish 
that our proposed CVI method either outperforms or is competitive with the other two VI approaches.
DGPs that correspond to the three logit models are considered, each using a small ($n=10,000$) and a large ($n=1$ or $5$ million) sample size. For the small samples, evaluation of the true posterior using MCMC is feasible, so that accuracy of the approximations can be assessed. It is 
difficult to evaluate the exact posterior for the large samples, so we compare the predictive accuracy of CVI
against that of AVI\footnote{Through out this paper, we do not use GPU computation or sub-sampling, which is different from the original paper of \cite{rodriguesScalingBayesianInference2022}.} and DAVI for these cases. %See Part~B of the Online Appendix for the details of the DAVI algorithm.
Table~\ref{tab:egsummary} provides a summary of the examples considered in this paper that we discuss below. 
\begin{sidewaystable}[p]
	\begin{center}
		\begin{threeparttable}
			\caption{Summary of the mixed effect examples}\label{tab:egsummary}
			{\small
				\begin{tabular}{lccccccccccc}
					\toprule
					& &\multicolumn{5}{c}{Example Features} &&\multicolumn{3}{c}{CVI Details} &\\ \cline{3-7} \cline{9-12}
					& \makecell{Model\\Specification} & $\mbox{dim}(\alphavec_i)$& $\mbox{dim}(\thetavec)$& $\mbox{dim}(\alphavec)$&$S$&\makecell{$n_{train}$} && $\mbox{dim}(\lambdavec)$& \makecell{Time/\\step (s)}&\makecell{Time to\\fit (min)} &\makecell{Benchmark\\ Methods}\\
					\hline
					\multicolumn{10}{l}{Simulated Examples}&&\\ 
					\cline{1-2}
					E.g. 1(a)	  & \makecell{MMNL} & 9&58&900&100&10,000&&391&0.0068&0.510&\makecell{AVI,\\ DAVI,\\MCMC}\\
					E.g. 1(b)	  & \makecell{MMNL} & 9&58&90,000&10,000&1,000,000&&391&0.2508&15.050&\makecell{AVI,\\ DAVI}\\
					
					E.g. 2(a)	  &  \makecell{B-MMNL}  & 9&62&900&100&10,000&&419&0.0133&0.91&\makecell{AVI,\\ DAVI,\\MCMC}\\
					E.g. 2(b)	  &  \makecell{B-MMNL}  & 9&62&90,000&10,000&1,000,000&&419&0.3487&29.06 &\makecell{AVI,\\ DAVI}\\
					
					E.g. 3(a)& \makecell{MNestL} & 9&60&900&100&10,000&&405&0.0211&0.98 &\makecell{DAVI,AVI}\\
					E.g. 3(b)&  \makecell{MNestL}  & 9&60&90,000&10,000&5,000,000&&405&1.7450&101.79&\makecell{DAVI,AVI}\\
					
					\hline
					\multicolumn{10}{l}{Consumer Choice Application}&&\\ 
					\cline{1-2}
					E.g. 4&  \makecell{MMNL} & 56&1,652&21,336&381&438,774&&11,554&0.9437&103.80&--\\
					
					E.g. 5&  \makecell{MNestL} & 56&1,656&21,336&381&438,774&&11,582&2.9051&150.10&--\\

					E.g. 6&  \makecell{B-MMNL} & 76&3,077&28,956&381&959,050&&21,529&6.1937&712.27&--\\			 					
					\bottomrule
				\end{tabular}
			}	
		\end{threeparttable}	
	\end{center}
	
	\begin{center}
		\begin{minipage}{0.85\linewidth} 	
			\centering
			\footnotesize
			\begin{itemize}[]
				\item Note: the example features include the dimension of the random coefficient vector $\alphavec_i$, global parameters $\thetavec$, latent variables $\alphavec$, number of groups $S$, and training data sample size $n$. Details of the CVI method include dimension of the variational parameters $\lambdavec$, and computation times are for a 2023 MacBook with Apple M3 Max processor. The final column indicates the benchmark or comparisons considered. We use $p=5$ as the number of factors in the factorized covariance matrix $V_0$ through out this paper.
			\end{itemize}
		\end{minipage} 		
	\end{center}
\end{sidewaystable}

\paragraph{Measuring heterogeneity:}
The purpose of fitting a mixed model is to account for heterogeneity. To assess this we compute the contribution of the random coefficients to the variance of the utility at~\eqref{v_itj_mmnl}
as follows. Let the design matrix $\xvec_{it} = (\xvec_{it2}^\top,\ldots,\xvec_{itJ}^\top)^\top$, %$\tilde{d}_i=\mbox{dim}(\alphavec_i)$, and $\tilde{d}_{ij}=\mbox{dim}(\alphavec_{ij})$. 
then a measure of total (TH) and alternative-specific (AH) contributions to heterogeneity (relative to the reference alternative) are: 

\begin{align}
	\text{TH}(\Sigma) 
	&= \frac{1}{n}\sum_{i,t}\mbox{Var}(\alphavec_{i}^\top \xvec_{it})
	= \frac{1}{n}\sum_{i,t}\xvec_{it}^\top \Sigma \xvec_{it}, 
	\label{TH} \\[1ex]
	\text{AH}_j(\Sigma_{(j-1)(j-1)}) 
	&= \frac{1}{n_j}\sum_{(i,t)|j\in C_{it}} \mbox{Var}(\alphavec_{ij}^\top \xvec_{itj})\nonumber \\
	&= \frac{1}{n_j}\sum_{(i,t)|j\in C_{it}}
	(\xvec_{itj})^\top 
	\Sigma_{(j-1)(j-1)}(\xvec_{itj}),
	\quad j=2,\ldots,J\,.
	\label{AH}
\end{align}
When computing $\text{AH}_j$ the summation is for all observations $(i,t)$ such that alternative $j$ is in choice set $C_{it}$, and $n_j$ is the number of such observations. Also recall that $\alphavec_{it1}-\bm{0}$ for identification.
These measures account for the varying choice sets and that the level of heterogeneity varies across $i,t$. When $\Sigma$ is known from the DGP the measures can be computed exactly. When using Bayesian inference, each measure can be computed using the variational
posterior mean of $\Sigma$. 

\paragraph{Predictions and predictive accuracy:} 
For any $(i,t)$ and covariate values $\xvec_{it1},\ldots,\xvec_{itJ}$, the predictive probability of selecting each alternative can be computed by Monte Carlo sampling from the the variational posterior. Using this mass function, the log-score and weighted $\mathbb{F}1$ score~\citep{sokolova2009systematic} can be computed  as measures of predictive accuracy. The $\mathbb{F}1$ score weights the binary $F_1$ scores for each alternative by their observed 
proportion ($\omega_j$) in the sample. Thus, 
$\mathbb{F}1= \sum_{j = 1}^{J} \omega_j F_{1,j}$,
where $F_{1,j}=2\text{TP}_j/(2\text{TP}_j+\text{FP}_j+\text{FN}_j)$, and $\text{TP}_j$, $\text{FP}_j$ and $\text{FN}_j$ are the true positives, false positives and false negatives for alternative $j$, respectively. See Part~C.2 of the Online Appendix for further details.

\subsection{Simulation 1: MMNL}\label{sec:sim_mmnl}
We generate data from the MMNL with $T=100$ observations per group for $S=100$ (small example)
and $S=10,000$ (large example) groups. {There are $J=4$ alternatives, four fixed-effect covariates ($w^f = 4$) and three random-effect covariates ($w^r = 3$) so that, with a reference category, $\dim(\alphavec_i)=9$;} full details of the DGP are found in Part~C of the Online Appendix.

\begin{figure}[htb]
	\centering
	\includegraphics[width=1\linewidth]{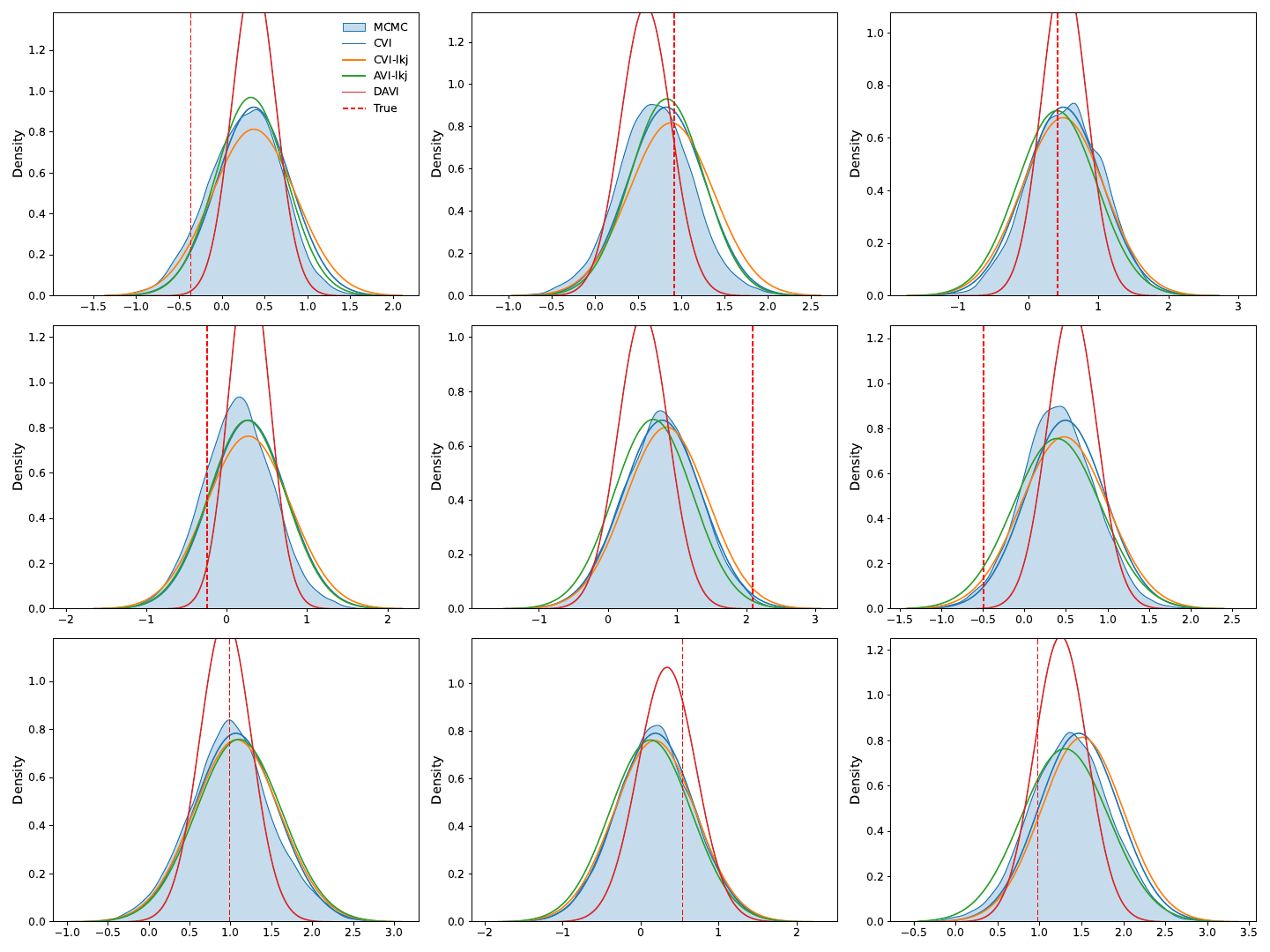}
	\caption{Posterior densities of random coefficients for a representative group in the small MMNL simulation. The exact posterior is shaded, while four variational posteriors are given as lines. Each row corresponds to random coefficients associated with a specific alternative, and each column corresponds to the random coefficients of a specific covariate. For identification purposes, the coefficients of the first (reference) alternative are fixed at $\zerovec$.}
	\label{fig:posterior_alpha_mmnl}
\end{figure}

We first illustrate our CVI method using the smaller example.
The approximated likelihood using CVI is relatively poor at the initial steps of the variational optimization algorithm, but improves significantly until around 300 steps, after which it remains stable. To illustrate this, profiles of the exact and approximated likelihood using CVI for a single representative group across optimization steps are given in Figure~\ref{fig:likelihood_approx_mmnl} in the Online Appendix. The resulting posterior approximation is very accurate. Figure~\ref{fig:posterior_alpha_mmnl} demonstrates this by plotting the variational posterior densities of the random coefficients for the same representative group.
Those 
obtained using CVI and AVI closely align with the exact posterior (computed using MCMC), while those from DAVI are less accurate. Here, CVI is fit using both the LKJ and HW priors for $\Sigma$, AVI with the LKJ prior as in~\cite{rodriguesScalingBayesianInference2022}, and DAVI with the HW prior; although results are insensitive to the choice of prior. 

\begin{table}[thb!]
	\captionsetup{skip=2pt}
	\centering
	\caption{Computational Performance of VI Methods in Simulations}
	\label{tab:results_comp}
	\footnotesize
	\begin{tabular}{lccccccccc}
		\toprule
		\midrule
		&&&&&&&&&\\
		\multicolumn{10}{c}{\textbf{Panel A: Simulation~1}} \\
		&&&&&&&&&\\\cline{2-10}
		&&&&&&&&&\\
			&
		& \multicolumn{4}{c}{E.g.~1(a): Small Example ($n=10,000$)} 
		& \multicolumn{4}{c}{E.g.~1(b): Large Example ($n=1,000,000$)} \\
		\cmidrule(lr){3-6} \cmidrule(lr){7-10}
		 
		\textbf{Method}& $p(\Sigma)$ &\makecell{Runtime\\ (min)}& \makecell{Total \\steps} & \makecell{Time/Step\\ (s)} & ELBO 
		 &\makecell{Runtime\\ (min)}&  \makecell{Total \\steps}& \makecell{Time/Step\\ (s)} & ELBO  \\ 
	%	\underline{\textbf{Method}}    							&&&&&&&&&\\
	%	&&&&&&&&&\\
		CVI &HW &0.85 & 4,500 & 0.0114 & {\bf -12,484}
						  & 33.35 & 5,900 & 0.3391 &  {\bf -1,215,605}\\
		DAVI &HW &1.03 & 8,200 & 0.0076 &  -12,586
		&42.05& 7,800 & 0.3235 &  -1,222,814\\								  
		CVI &LKJ &0.83 & 4,900 & 0.0101  & {\bf -12,486}
							&20.34 & 3,600 & 0.3391 &  {\bf -1,215,882}\\
		AVI &LKJ &0.89 & 7,600 & 0.0070  &-12,511
							&86.96 & 10,000 & 0.5218 &   -1,220,230\\
		\midrule\midrule
		&&&&&&&&&\\
		\multicolumn{10}{c}{\textbf{Panel B: Simulation~2}} \\			
		&&&&&&&&&\\\cline{2-10}
		&&&&&&&&&\\
		&
		& \multicolumn{4}{c}{E.g.~2(a): Small Example ($n=10,000$)} 
		& \multicolumn{4}{c}{E.g.~2(b): Large Example ($n=1,000,000$)} \\
		\cmidrule(lr){3-6} \cmidrule(lr){7-10}
		\textbf{Method}& $p(\Sigma)$ &\makecell{Runtime\\ (min)}& \makecell{Total \\steps} & \makecell{Time/Step\\ (s)} & ELBO 
&\makecell{Runtime\\ (min)}&  \makecell{Total \\steps}& \makecell{Time/Step\\ (s)} & ELBO  \\ 
%		\underline{\textbf{Method}}    							&&&&&&&&&\\
%		&&&&&&&&&\\
			{CVI} &HW& 1.32&4,100&0.0193&{\bf -18,503}
								& 24.86 & 3,600 & 0.4143&{\bf -1,873,250}\\
			{DAVI} &HW& 1.44&8,400&0.0103& -18,583
			& 26.86 & 4,400 & 0.3663&-1,883,099\\		
			{CVI}&LKJ &1.25&3,300&0.0226&{-18,522}
								&25.41 & 3,700 & 0.4121&{\bf -1,873,354}\\
			{AVI}&LKJ & 0.89&6,500&0.0082&{\bf -18,519} 
								&94.62 & 10,000 & 0.5677&-1,879,607\\
		\midrule\midrule
		&&&&&&&&&\\
		\multicolumn{10}{c}{\textbf{Panel C: Simulation~3}} \\
		&&&&&&&&&\\\cline{2-10}
		&&&&&&&&&\\
		&
		& \multicolumn{4}{c}{E.g.~3(a): Small Example ($n=10,000$)} 
		& \multicolumn{4}{c}{E.g.~3(b): Large Example ($n=5,000,000$)} \\
		\cmidrule(lr){3-6} \cmidrule(lr){7-10}
		 
		\textbf{Method}& $p(\Sigma)$ &\makecell{Runtime\\ (min)}& \makecell{Total \\steps} & \makecell{Time/Step\\ (s)} & ELBO 
&\makecell{Runtime\\ (min)}&  \makecell{Total \\steps}& \makecell{Time/Step\\ (s)} & ELBO  \\ 

%		\underline{\textbf{Method}}    							&&&&&&&&&\\
%		&&&&&&&&&\\
		CVI &HW&0.98&2,800&0.0211& {\bf -8,972}
						&101.79&3,500&1.7450&{\bf -4,256,502}\\
		DAVI &HW&2.29&6,200&0.0222& -8,972 
	&304.82&5,600& 3.2659&-4,275,165\\						
		CVI &LKJ&1.17&3,000&0.0233&-8,970
						&108.17&3,700&1.7541&{\bf -4,256,068}\\
		AVI& LKJ&1.32&9,300&0.0085&{\bf -8,925}
						&469.52& 10,000& 2.8174&-4,287,867\\
		\bottomrule
	\end{tabular}
	
	\vspace{0.75ex}
	\begin{minipage}{0.95\linewidth}
		\footnotesize
	Note: Computation times are for a 2023 MacBook with Apple M3 Max processor. Higher ELBO values correspond to greater accuracy, with the maximum value for each example/prior $p(\Sigma)$ combination in bold. Total runtime is for the stopping rule outlined in~\cite{rodriguesScalingBayesianInference2022}.
	\end{minipage}
\end{table}

Panel~A in Table~\ref{tab:results_comp}
compares the computational demand and accuracy of the different methods for both the small and large datasets. Accuracy of the approximation is measured using the ELBO values, and in both examples and for both priors CVI is more accurate.
Computational time is similar for all three methods in the small example, but CVI is between 107\% and 328\% faster than the DAVI and AVI methods in the large example where speed is more important. Choice of prior has little impact, a result we find in all three simulations. 
Table~\ref{tab:MMNL_sim_TH} reports the heterogeneity estimates for the different methods, 
along with the true values for the DGP.
For the small example, the estimates obtained using the exact posterior are also reported. All methods do a good job at capturing the level of heterogeneity, with CVI and AVI close to the exact posterior estimates, and DAVI under-estimating it slightly. Results from the large example are more accurate, which is due to an improvement in 
the estimates of the random coefficients covariance matrix $\Sigma$ in this case. 

\begin{table}[htbp]
	\begin{center}
		\caption{Estimates of Heterogeneity in Simulation~1}
		\label{tab:MMNL_sim_TH}
		\begin{tabular}{cccccccccc}
			\toprule
			&
			& \multicolumn{4}{c}{E.g.~1(a): Small Example} 
			& \multicolumn{4}{c}{E.g.~1(b): Large Example} \\
			\cmidrule(lr){3-6} \cmidrule(lr){7-10}
			Method 
		& $p(\Sigma)$ & TH & AH(2) & AH(3) & AH(4) 
			& TH & AH(2) & AH(3) & AH(4) \\ \hline
			True    
			& {--} 
			& 23.464 & 1.356 & 1.572 & 1.484 
			& 11.424 & 1.264 & 1.424 & 0.988 \\
			MCMC
			& {HW} 
			& 20.664 & 1.132  & 1.596  & 1.372
			& -- & -- & -- & -- \\
			CVI     
			& {HW} 
			& 18.088 & 0.972 & 1.380 & 1.236 
			& 10.976 & 1.156 & 1.300 & 0.936 \\
			CVI
			& {LKJ} 
			& 20.664 & 1.176 & 1.576 & 1.352 
			& 10.864 & 1.148 & 1.304 & 0.948 \\
			AVI
			& {LKJ} 
			& 20.888 & 1.132 & 1.600 & 1.404 
			& 10.472 & 1.164 & 1.312 & 0.912 \\
			DAVI    
			& {HW} 
			& 15.344 & 0.852 & 1.260 & 1.068 
			& 9.856 & 1.180 & 1.220 & 0.908 \\
			\bottomrule
		\end{tabular}
	\end{center}
	Note: Values reported for the DGP are labeled ``True'', exact and variational posterior mean values (computed using Monte Carlo simulation as discussed in the text) are reported for the estimators.
\end{table}

\begin{figure}
	\centering
	\includegraphics[width=0.85\linewidth]{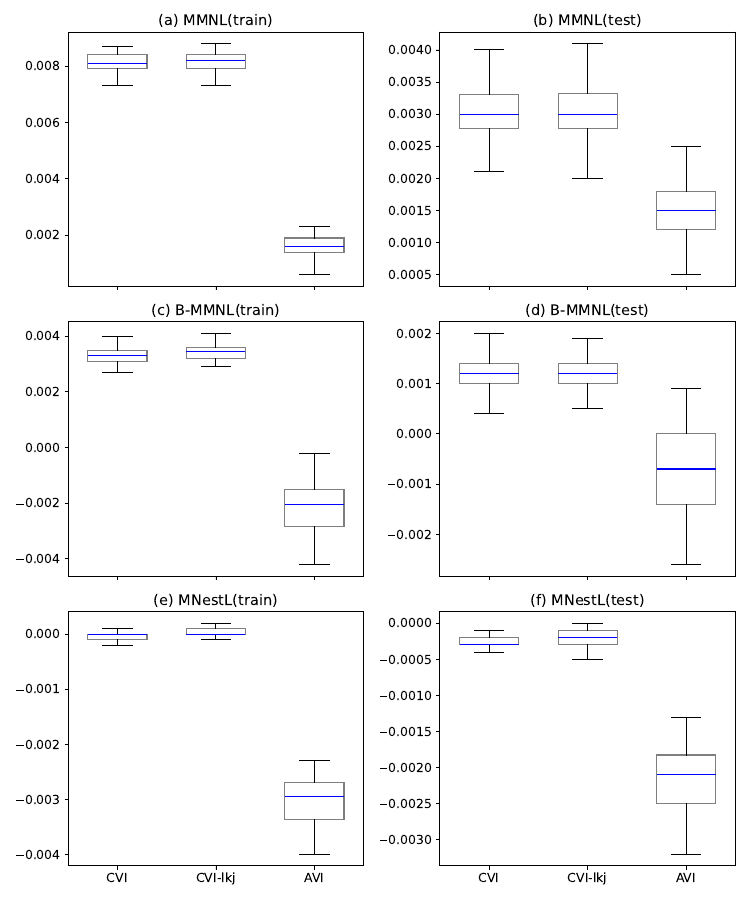}
	\caption{Comparison of $\mathbb{F}$1 scores from four VI methods for the large simulation examples. Values are the differences in scores between each VI method and DAVI. The rows correspond to Simulations 1, 2 and 3. The first column shows scores for the training data, and the second column shows scores for the test data. 
	Positive/negative values indicate higher/lower predictive accuracy than DAVI. Boxplots exclude outliers, defined as observations more than 1.5 × IQR from the box. Equivalent plots for the log-score are found in Figure~\ref{fig:robust_ls} of the Online Appendix.}
	\label{fig:predictF1}
\end{figure}

To compare predictive accuracy we generate 100 replicate training datasets from the large example (i.e E.g. 1(b)). For each replicate a further 30 observations from each of the same $S=10,000$ groups were generated as test data. Figure~\ref{fig:predictF1} gives boxplots of the $\mathbb{F}1$ score for predictions for the train (panel~a) and test (panel~b) data fit using the different methods.
The scores are reported for CVI and AVI as the difference with DAVI, so that positive/negative values indicate greater/lower accuracy than DAVI. 
Both CVI and AVI produce more accurate predictions than DAVI, with those computed using CVI most accurate.

\subsection{Simulation 2: B-MMNL} \label{sec: sim_mixbc}
We repeat Simulation~1, but allowing for bundles using the B-MMNL model. As before, there are four
singleton alternatives (labeled A,B,C,D), but now include the four bundles \{A,C\}, \{A,D\}, 
\{B,C\} and \{B,D\}, giving
a total of $R=8$ alternatives. In~\eqref{v_itj_mixbc}, $\gamma_1=\ldots=\gamma_4= 0$, $\gamma_5=0.0976$, $\gamma_6=0.4304$, $\gamma_7=0.2055$ and $\gamma_8=0.0898$. 

In this example, CVI and AVI estimate the posteriors of the random coefficients well, while those from DAVI are more biased and under-estimate posterior uncertainty; 
see Figure~\ref{fig:posterior_alpha_mixbc} in the Online Appendix. Panel~B in Table~\ref{tab:results_comp} reports the performance of the methods, with higher ELBO values for CVI and AVI indicating increased accuracy. 
The computational cost of the VI methods are comparable in the small example, while CVI is 272\% faster than AVI for the large example. 
Heterogeneity estimates are reported in Table~\ref{tab:Mixbc_sim_TH} of the Online Appendix, with results similar to those in Simulation~1, with the exception that AVI substantially under-states heterogeneity. As before, CVI heterogeneity estimates are more accurate than those from the other VI methods. All VI methods give accurate estimates of the complementary parameters $\gamma_5,\ldots,\gamma_8$; see Table~\ref{tab:Mixbc_gamma_sim} of the Online Appendix. 

Finally, we repeat the prediction exercise in Simulation~1 by generating 100 replicates of the training data for the large example (E.g.~2(b)), and additional test data as before. 
Figure~\ref{fig:predictF1}(c,d) shows that predictions from the model fitted using CVI out-perform the other VI methods as measured using the $\mathbb{F}$1 score, with  
the same conclusion being drawn when using the log-score in Figure~\ref{fig:robust_ls} of the Online Appendix.

\subsection{Simulation~3: MNestL}\label{Sec:simulNest}
We extend the simulation to the MNestL model with $J=4$ alternatives divided into $K=2$ nests, with the correlation between each nest $\tau_1=0.3$ and $\tau_2=0.7$. For the large example, the number of observations per group in the training data is increased to 500, making this the largest example in the paper with $n=5$ million total observations; all other simulation settings remain unchanged. The likelihood of a MNestL model has a more complex geometry than those of MMNL and B-MMNL, and exact posterior results were difficult to evaluate using MCMC and are not reported for the small example. 

The likelihood function is not log-concave everywhere \citep{trainDiscreteChoiceMethods2009}, so that the matrix $-\nabla^2_a l(\thetavec,\avec_i)$ can be negative definite. Therefore,
to construct the approximation in CVI we follow \cite{highampsd1988} and compute the nearest symmetric positive semi-definite matrix to \( -\nabla^2_a l(\thetavec,\avec_i) \), that ensure $V_i$ is always positive definite as follows.

Consider a symmetric $q\times q$ matrix $A$ with spectral decomposition $A = Q\Lambda Q^\top$, where $Q$ is the eigenvector matrix, and $\Lambda$ is the diagonal matrix of the eigenvalues of $A$. Then a positive semi-definite approximation of $A$ is 
\begin{equation}
\tilde{A}   = Q	\widetilde{\Lambda} Q^\top \label{pdf2},
\end{equation}
where the diagonal matrix $\widetilde{\Lambda} = \text{diag}( \widetilde{\Lambda}_1, \dots,  \widetilde{\Lambda}_q)$ with elements $\widetilde{\Lambda_i}  = \Lambda_i$ if $\Lambda_i \geq 0$ and zero otherwise. For the MNestL model, we first calculate $-\nabla^2_al(\thetavec,\avec_i)$ and then replace it with its nearest symmetric positive semi-definite matrix~\eqref{pdf2} when computing $V_i$ in CVI.

Panel~C of Table~\ref{tab:results_comp} reports the performance of the VI methods, with CVI again much faster than DAVI and AVI for the large example, and producing a more accurate variational approximation of the posterior as measured by the ELBO function. 
However, CVI tends to under-estimate heterogeneity by a factor of around 29\%, whereas AVI and DAVI capture the heterogeneity well; see Table~\ref{tab:MixNL_sim_TH} in the Online Appendix. 

Finally, we generate 50 replicates of the large dataset in Example~3(b) as training data, and then use these to predict both the training data and a further 50 test datasets. In this simulation we use 50, rather than 100 replicates as in Simulations~1 and~2, because of the long time taken by AVI and DAVI to fit each dataset. Figure~\ref{fig:predictF1}(e,f) produce the equivalent boxplots for the MNestL, and it can be seen AVI produces less
accurate predictions that either DAVI and CVI. Unlike the other two simulations, DAVI produces a fitted model that gives slightly more accurate predictions than CVI.  

\subsection{Summary of Simulation Findings}
In summary, all three VI methods are effective approaches to compute posterior inference for large mixed multinomial logit models. This includes more complex variants including the MNestL, for which likelihood-based inference is difficult to compute due to its complex geometry. For both the MMNL and B-MMNL models, CVI is clearly the most accurate VI method, and is competitive with DAVI for the MNestL model. 
Our proposed CVI method is also substantially faster than the two alternatives, making it an attractive choice for large datasets. For example, it only takes 20-30mins to fit an MMNL with $S=10,000$ groups and 1 million observations, and around 100mins to fit an MNestL with $S=10,000$ groups and 5 million observations, using a standard laptop.

\section{Large scale consumer choice application}\label{sec: empirical}
We use our methodology to model purchases of pasta by households across different stores of a U.S. grocery chain using a large scanner panel dataset. When modeling individual purchases, there is likely to be extensive variability at the store and product levels in the drivers of consumer choice~\citep{hochDeterminantsStoreLevelPrice1995,dellavignaUniformPricingUS2019,hitschPricesPromotionsUS2021}. 
Our objective is to use the three variants of the mixed multinomial logit to capture this heterogeneity for price and two promotion variables, and to measure its impact on predictive accuracy. By doing so, we also use the store level estimates to determine some key store
characteristics that drive variation in price elasticities. Finally, we explore if including information on purchases of pasta sauce in the mixed bundle choice model improves the modeling of pasta sales. 

However, computing inference for these mixed models 
is challenging for three reasons. First, the dataset contains over half a million choice records. Second, there are $15$ main pasta alternatives and $4$ alternative-specific covariates in the data,
resulting in up to $56$ random coefficients per store in the MMNL and MNestL models. When considering bundling with only $5$ brands of pasta sauces, this further increases to $76$ random coefficients per store in the B-MMNL model. This makes application of likelihood-based inference, including MCMC methods, difficult computationally. Third, the panel is unbalanced, with the number of transactions per store in the training sample ranging from $53$ to $3,260$, making it difficult to apply AVI. However, the proposed CVI method can be used to compute inference at scale, with an unrestricted $\Sigma$ matrix, as we do here for this application.

\subsection{Pasta Data}
 The dataset, referred to as the Carbo-loading data, consists of transactions sourced from the Dunnhumby data platform. It records consumer purchases of pasta and pasta sauce products over a two-year period (104 weeks) at a U.S. grocery chain.\footnote{\url{https://www.dunnhumby.com/source-files/}}
 We exclude all transactions prior to week 43, as promotion data are unavailable for earlier weeks. We focus on sales of pasta from the four most popular brands in the dataset (Private Label, Barilla, Mueller and Creamette) and then the four most popular pasta types within this group (thin spaghetti, elbow macaroni, spaghetti and angel hair). This yields a total of 15 alternatives because the combination of angel hair pasta and the Creamette brand does not appear in the data. The resulting dataset contains 548,647 observations across 381 stores. We choose the most frequently purchased alternative ``Private Label thin spaghetti'' as the reference case. Inclusion of data on pasta sauce is discussed later in Section~\ref{sec: empirical_mixbc}.

Let $y_{it} \in C_{it}$ denote the product chosen in transaction $t$ at store $i$, where $C_{it}$ is a choice set that varies by $t$ and $i$. 
The index $t$ spans all transactions over the period by all households. Each pasta is uniquely defined by the two attributes of brand and type;
for example, ``Barilla thin spaghetti'' is a valid alternative, whereas ``thin spaghetti'' is not because it does not uniquely identify a specific product. Table~\ref{tab:pasta_summary} summarizes the availability of the different alternatives across transaction instances (i.e. across choice sets), along with market share in our dataset.

\begin{table}[thb!]
	\captionsetup{skip=2pt}
\begin{center}
	\caption{Availability and Market Share of Pasta Alternatives.}
	\label{tab:pasta_summary}
{\small 
	\begin{tabular}{lcccccccccc}
		\toprule
		& \multicolumn{4}{c}{\textbf{A: Availability (\%)}} &  & \multicolumn{4}{c}{\textbf{B: Market Shares (\%)}} &  \\
		&      \multicolumn{4}{c}{\textbf{Pasta Brand}}       &  &       \multicolumn{4}{c}{\textbf{Pasta Brand}}       &  \\ \cline{2-5}\cline{7-10}
		&        &        &        &                          &  &        &        &        &                           &  \\
		& Prv-Lab &  Barilla   &  Mueller   &           Creamette           &  & Prv-Lab &  Barilla   &  Mueller   &            Creamette            &  \\
		\textbf{Pasta Type} &        &        &        &                          &  &        &        &        &                           &  \\
		Thin Spaghetti      &   100    & 54.82 & 37.43 &          43.71          &  &   22.62   & 3.27 & 2.46 &          3.14           &  \\
		Spaghetti           & 98.36 & 56.82 & 43.12 &          47.40          &  & 19.82 & 3.46 & 3.22 &          3.96           &  \\
		Macaroni            & 94.89 & 36.71& 40.65 &          46.16          &  & 15.03 & 2.01 & 2.88&          3.65           &  \\
		Angel Hair          & 87.84 & 54.26 & 36.72 &            0             &  & 9.14& 3.31& 2.04 &             0             &  \\ \bottomrule
	\end{tabular}
}
\end{center}
	\small 
		Panel~A reports availability of alternatives across choice sets (in \%), while Panel~B reports market share of alternatives measured by sale incidence (in \%). Results are reported segmented by pasta brand in columns and pasta type in rows.
\end{table}

There are three covariates in the data, with the first being the logarithm of the purchase price per ounce (\textit{lnprice}). The other two are promotion dummy variables indicating whether or not the product was on an in-store display (\textit{display}) or advertised in the weekly mail circular (\textit{feature}). The covariates vary by store $i$, transaction $t$, and alternative $j$, so that $\xvec^r_{itj} = \left( 1, {lnprice}_{itj}, {display}_{itj}, {feature}_{itj} \right)^\top$ including the intercept. Our objective is to model the heterogeneity across stores and alternatives, represented by the random coefficient vector $\boldsymbol{\alpha}_{ij}$ in~\eqref{v_itj_mmnl}.

The covariate values are only recorded for alternatives that are purchased, and to impute their values for the other alternatives in each transaction we apply the following procedure. 
For \textit{lnprice}, we use the average price of the same alternative at the same store on the same day; if still missing, we use the average at the same store during the same week. If price remains missing after this step, that alternative is removed from the choice set $C_{it}$ as it is likely to be unavailable at instance $t$ in store $i$. 
For \textit{display} (and analogously for \textit{feature}), we instead use the maximum available value of the same alternative at the same store on the same day; if still missing, we use the maximum during the same week. If display is still missing at this point, we set it to zero.

To assess predictive performance, we randomly partition the transactions within each store into a training set (80\%) and a held-out test set (20\%), so that there are 438,774 observations in the training data and 109,873 observations in the testing data. We employ two benchmarks for predictive purposes. The first is a
na\"ive approach that uses the empirical in-sample marginal choice probabilities as a predictive distribution.
The second is comprised of the same multinomial logit models (i.e. standard, mixed and bundle) but with fixed coefficients across stores (but varying by alternative) estimated using stochastic variational inference.

\subsection{Pasta Choice}\label{sec: empirical_pasta}
We use the proposed CVI method to estimate both MMNL and MNestL models using the training data.
For the MNestL model, we group the choice alternatives into the following four nests that correspond to first selecting pasta type, followed by brand:
\begin{itemize}
	\item $N_1 =$\{thin spag Private Label; thin spag Barilla; thin spag Mueller; thin spag Creamette\}
	\item $N_2 =$\{spaghetti Private Label; spaghetti Barilla; spaghetti Mueller; spaghetti Creamette\}
	\item $N_3 =$\{macaroni Private Label; macaroni Barilla; macaroni Mueller; macaroni Creamette\}
	\item $N_4 =$\{angel hair Private Label; angel hair Barilla; angel hair Mueller\}.
\end{itemize} 
The assumption of varying choice sets then ensures that $B_{itk}\subseteq N_k$, for $k \in \{1,2,3,4\}$ in~\eqref{p_mixnl}, depending on availability of alternatives. 

Table~\ref{tab:comp_models_pasta} summarizes the computational and predictive performances for the MMNL and MNestL models. In terms of runtime, the MMNL model converges faster than the MNestL model, due to the lower computational cost per step of the optimization algorithm. Predictions are evaluated in the same manner as in Section~\ref{sec:sim}, and their accuracy measured using the $\mathbb{F}1$ scores for both the training and test data. The results from the MMNL and MNestL models are very similar, suggesting that the incorporation of sequential decision-making of pasta type, followed by brand,  does not translate into more accurate predictions. The fixed coefficients benchmark models (MNL and NestL) are much faster to fit, but are substantially less accurate, highlighting the importance of modeling heterogeneity across stores.

\begin{table}[thb!]
	\captionsetup{skip=2pt}
\begin{center}
	\caption{Computational Speed and Predictive Accuracy for Consumer Choice Application}
	\label{tab:comp_models_pasta}
	\begin{tabular}{lccccc}
		\toprule\midrule
		                            & Runtime (min) & Total steps & Time/step (s) &        \multicolumn{2}{c}{$\mathbb{F}1$}\\		\cline{5-6}
\underline{\textbf{Model}} &               &             &               & Train& Test \\
		 Na\"ive                         &   --     &  --   &    --     &          0.0830                    &        0.0855    \\                           
		MMNL                         &    103.80     &   6{,}600   &    0.9437     &          {0.1766}                      &        {0.1745}    \\
		MNL                      &     10.72     &   4{,}200   &    0.1531     &               0.1345                          &   0.1371              \\
		MNestL                         &    150.10     &   3{,}100   &    2.9051     &           {0.1771}             &         {0.1749}          \\
		NestL                      &     31.42     &   3{,}900   &    0.4835     &                      0.1378                       &  0.1405 \\
B-MMNL                         &     712.27     &   6{,}900   &    6.1937      &                 {0.1802}          &         {0.1781}          \\
B-MNL                      &     96.86    &   4{,}700   &    1.2365      &      0.1288           &         0.1314         \\
 \bottomrule
	\end{tabular}\\
	\end{center}
	\small
     MNL, NestL and B-MNL denote the fixed effect multinomial logit, nested logit and bundle choice models, respectively. For the bundle models, the predictions for pasta choice are for the same transactions obtained by marginalizing over the sauce alternatives, so their accuracy is directly comparable to that of the other models. Higher $\mathbb{F}1$ values indicate greater accuracy.
\end{table}

\begin{table}[thb!]
	\captionsetup{skip=2pt}
	\centering
	\caption{Heterogeneity Estimates}
	\label{tab:heterogeneity_pasta_final_clean}
	\footnotesize
	
	% =======================
	% Panel A: MMNL (4x4 x 2) — Thin×PriLab corrected
	% =======================
	\begin{tabular}{@{\hspace{4\tabcolsep}}lccccc@{\hspace{4\tabcolsep}}}
		\midrule\midrule
		                    &                             \multicolumn{5}{c}{\textbf{Panel A: MMNL \ \ \ \ \ \ }}                             \\
		                      &       \multicolumn{4}{c}{\textbf{Pasta brand}}       &  \\ \cline{2-5}
		                    &        &        &        &                          &    \\
		                      & Prv-Lab &  Barilla   &  Mueller   &           Creamette             &  \\
		\textbf{Pasta type}   &        &        &        &                           &  \\
		Thin Spaghetti        & --    & 12.08\% & 19.83\% & 14.12\% &  \\
		Spaghetti             & 6.47\% & 11.91\% & 18.55\% & 11.71\%&  \\
		Macaroni              &7.15\% & 12.01 \%& 17.83\% & 11.77\%&  \\
		Angel Hair            & 7.06\% & 13.88\%& 14.11 \%& --&  \\ \midrule\midrule
	\end{tabular}
	
	\vspace{1.0ex}
	% =======================
% Panel B: MixNL (4x4 x 2) — Thin×PriLab corrected
% =======================
\begin{tabular}{@{\hspace{4\tabcolsep}}lccccc@{\hspace{4\tabcolsep}}}
	&                            \multicolumn{5}{c}{\textbf{Panel B: MNestL \ \ \ \ \ \ }}                             \\
  &       \multicolumn{4}{c}{\textbf{Pasta brand}}       &  \\ \cline{2-5}
	  &        &        &        &                           &  \\
	 & Prv-Lab &  Barilla   &  Mueller   &           Creamette       &  \\
		\textbf{Pasta type}   &        &        &        &                           &  \\
		Thin Spaghetti        & --    & 16.85\% & 26.37\% & 18.40\% &  \\
Spaghetti             & 6.98\% & 14.17\% & 21.72\% & 13.63\%&  \\
Macaroni              &7.37\% & 14.60 \%& 21.23\% & 13.35\%&  \\
Angel Hair            & 7.70\% & 14.64\%& 15.05 \%& --&  \\ \midrule\midrule
\end{tabular}	
	
	\vspace{1.0ex}

% =======================
% Panel C: MixMC (category-native labels; first column is PriLab for pasta brand)
% =======================
\begin{tabular}{@{\hspace{4\tabcolsep}}lcccccc}
	&                            \multicolumn{6}{c}{\textbf{Panel C: B-MMNL  \ \ \ \ \ \ }}                             \\
	&       \multicolumn{4}{c}{\textbf{Pasta brand}}  &     &  \\ \cline{2-5}
	&        &        &        &             &              &  \\
	& Prv-Lab &  Barilla   &  Mueller   &           Creamette       &  &\\
	\textbf{Pasta type}   &        &        &        &                           &  &\\
	Thin Spaghetti        & --    & 11.65\%& 23.04\%& 32.16\% & & \\
	Spaghetti             & 6.72\%& 11.20\%&20.83\%&28.41\%&  &\\
	Macaroni              &12.98\%&15.61\%&21.29\%&29.6\%& & \\
	Angel Hair            & 6.75\%&13.30\%&25.00\%& --&  &\\ \midrule
	& Ragu & Prego & Prv-Lab & Hunt`s&Mixed &\\
	Sauce Brand& 6.17\%&  9.87\%& 13.02\%& 10.02\%& 16.87&\\
	\bottomrule
\end{tabular}	
	
	\vspace{0.75ex}
	\begin{minipage}{0.65\linewidth}
		\footnotesize
		Panels A and B report $\calR_j$ as defined in (\ref{R_j}) by pasta type (rows) and brand (columns). Panel~C reports $\calR_j$ for both pasta and sauce groups. 
	\end{minipage}
\end{table}

We measure the heterogeneity of each alternative for both fitted models as follows. First, for each 
alternative $j=2,\ldots,J$, we  compute the estimated contribution of the random coefficients to the utility variance $AH_j$ as in~\eqref{AH}. Then, in Table~\ref{tab:heterogeneity_pasta_final_clean} we report these values as a proportion of the variation in the total utility for alternative $j$:
\begin{align}
	\calR_j = \frac{\text{AH}_j}{\text{AH}_j+\pi^2/6}\label{R_j},
\end{align}
where $\mbox{Var}(\varepsilon_{itj})=\pi^2/6$ is the marginal variance of the disturbance $\varepsilon_{itj}$ for all three logistic models;  see~\citet[p.35]{trainDiscreteChoiceMethods2009}.% and~\cite{heissStructuralChoiceAnalysis2002}.
The heterogeneity is substantial for both the MMNL and MNestL models, which have similar results. Moreover,
across pasta types the magnitude of heterogeneity remains similar, but differs over brand, suggesting that store-based heterogeneity is primarily in the response to product brand.

With the fixed coefficient models, Table~\ref{tab:comp_models_pasta} 
shows that NestL has more accurate predictions than MNL. 
This difference is not observed with the mixed models,
with the MMNL and MNestL models having very similar predictive accuracy and heterogeneity estimates. This is because 
the MMNL model can approximate a wide range of substitution patterns through the correlated random coefficients and often provides a good approximation to MNestL~\citep{mcfaddenMixedMNLModels2000}. 

\subsection{Covariate Effects}
We now analyze the effect of the covariates for the MMNL model in three different ways.\footnote{The results are very similar for the MNestL model, so that we do not present these here.} The first is measuring the size of store-based heterogeneity associated with each covariate. 
Let $k = 1,\dots,4$ index the covariates including the intercept (i.e. $1$, \textit{lnprice}, \textit{display}, \textit{feature}), and $\Sigma^{(k)}_{it}$ denote the $J_{it}\times J_{it}$ submatrix of $\Sigma$ that corresponds to covariate $k$ for the $J_{it}\equiv|C_{it}|$ alternatives at observation $(i,t)$. Let $\xvec_{it}^{(k)}$ and $\alphavec_{i}^{(k)}$ stack the values of the $k$th covariate and random coefficient, respectively, across these $J_{it}$ alternatives. Similar to the total and alternative heterogeneity measures, we evaluate the variance associated with the random coefficients for covariate $k$ as
\begin{align*}
	\text{CH}_k(\Sigma)={\frac{1}{n}}\sum_{i,t}\frac{1}{J_{it}}\mbox{Var}((\alphavec^{(k)}_{i})^\top \xvec_{it}^{(k)}) = {\frac{1}{n}}
	\sum_{i,t}\frac{1}{J_{it}}
	\xvec_{it}^{(k)\top}
	\Sigma^{(k)}_{it}\,
	\xvec_{it}^{(k)}\,.
\end{align*} 
This is estimated at the posterior mean estimate of $\Sigma$, and Table~\ref{tab:cov_heterogeneity} reports the results. Most variability is associated with the intercept, which reflects the high heterogeneity associated with the alternatives themselves. For the observed covariates, the greatest source of 
heterogeneity is due to price, which is unsurprising as pasta is a low value, low
involvement and price sensitive good.

\begin{table}[ht!]
	\begin{center}
	\caption{Covariate-specific Heterogeneity Measures}
	\label{tab:cov_heterogeneity}
	\begin{tabular}{lcccc}
		\hline
		& Intercept   &\textit{lnprice }    & \textit{display }    &\textit{feature  }   \\
		$\text{CH}_k(\Sigma)$&0.588&	0.056& 0.027  &0.017  \\
		\hline
	\end{tabular}\end{center}
\end{table}

\begin{figure}[thb]
	\centering
\includegraphics[scale= 0.65]{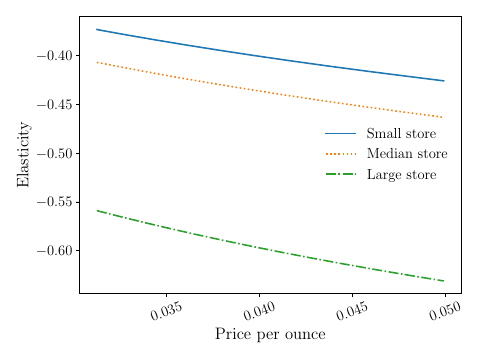}	
\caption{Own-price elasticities in for the most popular pasta alternative ``Private Label Thin Spaghetti''. They are computed for three representative stores using their store-specific random coefficients.}
\label{Fig:elasticities}
\end{figure}

The second analysis is to compute store-specific own price elasticities. 
Following~\cite{greeneDoesScaleHeterogeneity2010}, these are nonlinear functions of all the covariates and we report the elasticities evaluated at the sample means of the other covariates; see Part~E.1 in the Online Appendix for details.
Figure~\ref{Fig:elasticities} plots the elasticities of the ``Private Label thin spaghetti'' alternative, evaluated using the random coefficients of three representative stores selected according to size (the $25^{th}$, $50^{th}$ and $75^{th}$ size percentiles). 
The results indicate that customers of the smaller store are less price elastic for this alternative, whereas those of the larger store are more price elastic-- possibly reflecting greater assortment and substitution opportunities in larger stores. 
Across all three stores the elasticities become more negative as price rises, indicating that consumers are generally more price sensitive at higher prices. 

The third analysis is to profile the price
elasticities across store in more detail using the following three store characteristics:
\begin{itemize}
\item Geography: a dummy variable for the two regions recorded in the database, with ``Area 1'' coded as 0, and ``Area 2'' coded as 1.
\item Store Size:  logarithm of the total number of transactions at each store.
\item Store Premium: logarithm of the average sale price per transaction at each store. 
\end{itemize}
For each pasta alternative, Table~\ref{tab:regression_elasticity} presents the coefficients of regressions with elasticity as the dependent variable, and geography, store size and premium sales as covariates. Four observations can be drawn.
First, the results are consistent across products, with almost all store characteristics significant. 
Second, stores located in Area 2 exhibit more negative elasticities (i.e. higher price sensitivity) than stores in Area 1. Third, larger stores have more negative elasticities, indicating that customers of larger stores tend to be more price sensitive. Fourth, stores with higher premiums have smaller elasticities (i.e. less negative) for most alternatives, suggesting that customers in premium stores are less price sensitive than those of value stores.

\begin{table}[htbp]
	\begin{center}
	\caption{Regressions of Own-Price Elasticities on Store Characteristics}
	\label{tab:regression_elasticity}
	\begin{tabular}{lccc}
		\hline
		\textbf{Pasta Alternative} & \textbf{Geography} & \textbf{Store Size} & \textbf{Premium Sales} \\
		Thin--PriLab & -0.043 (0.008) & -0.090 (0.008) & 0.299 (0.066) \\
		Thin--Bar    & -0.072 (0.018) & -0.121 (0.020) & 0.416 (0.157) \\
		Thin--Mue    & -0.049 (0.017) & -0.117 (0.019) & 0.525 (0.150) \\
		Thin--Cre    & -0.131 (0.019) & -0.167 (0.021) & 0.400 (0.167) \\
		\hline
		Spa--PriLab  & -0.032 (0.014) & -0.097 (0.016) & 0.303 (0.124) \\
		Spa--Bar     & -0.071 (0.016) & -0.124 (0.018) & 0.008 (0.144) \\
		Spa--Mue     & -0.131 (0.015) & -0.159 (0.016) & 0.801 (0.130) \\
		Spa--Cre     & -0.084 (0.016) & -0.138 (0.018) & 0.534 (0.142) \\
		\hline
		Mac--PriLab  & -0.019 (0.013) & -0.119 (0.014) & 0.523 (0.114) \\
		Mac--Bar     & -0.174 (0.014) & -0.139 (0.016) & 0.520 (0.124) \\
		Mac--Mue     & -0.002 (0.015) & -0.076 (0.016) & 0.537 (0.128) \\
		Mac--Cre     & -0.109 (0.014) & -0.107 (0.015) & 0.403 (0.119) \\
		\hline
		Ang--PriLab  & -0.092 (0.013) & -0.069 (0.015) & -0.244 (0.118) \\
		Ang--Bar     & -0.060 (0.011) & -0.088 (0.012) & 0.487 (0.099) \\
		Ang--Cre     & -0.137 (0.011) & -0.102 (0.012) & 0.020 (0.096) \\
		\hline
	\end{tabular}
	\end{center}
		Note: The table presents estimated coefficients with standard errors in  parenthesis. 
\end{table}

 \subsection{Pasta and Sauce Choice}\label{sec: empirical_mixbc}
To fit the B-MMNL we expand the dataset to include purchases of pasta sauce. We consider bundles of two alternatives, where one is a pasta and the other is a pasta sauce. There are five alternative sauces, including the four top brands ``Ragu'', ``Prego'', ``Private Label'' and ``Hunt's'', which together account for over 80\% of sales. The fifth alternative is labeled ``Mixed'' which includes all other brands plus instances of multi-brand purchases. For example, if a basket has two Ragu sauces, it is coded as a ``Ragu'' purchase, whereas if there is one Ragu and Prego, then it is coded as ``Mixed''. Further details on data construction are given in Part~D of the Online Appendix. Overall, there are $J = 20$ unique single product alternatives and $R = 95$ possible bundles, including singleton bundles consisting of only one product.

The final dataset consist of 1,199,242 transactions across the $381$ stores. The predictor set is identical to that used in the MMNL model. We retain the same training–testing partition for transactions involving pasta (including both pasta singletons and bundles) as in Section~\ref{sec: empirical_pasta}, allowing direct comparison with the MMNL and MNestL results. We randomly partition the remaining data (pasta sauce only transactions) in each store into the training (80\%) and testing (20\%) data, so that the train and test data contain 959,050 and 240,192 observations, respectively. Again, we consider random coefficients for the full set of predictors with {an unrestricted} $\Sigma$ matrix, making estimation a challenging task. 

The last two rows of Table~\ref{tab:comp_models_pasta} summarize the results for the B-MMNL and the fixed coefficient equivalent B-MNL model.  
We evaluate predictive performance for pasta by marginalizing over the sauce alternatives, enabling direct comparison with the other models; see Part~E.2 of the Online Appendix for details.

The 
B-MMNL model more accurately predicts pasta choice compared to the MMNL and MNestL models, suggesting that differentiating between single and bundled choices provides useful information on the choice of pasta alone. The B-MMNL model also outperforms the fixed effect model B-MNL substantially, re-confirming the importance of heterogeneity. 
Panel~C of Table~\ref{tab:heterogeneity_pasta_final_clean} reports $\calR_j$ heterogeneity metrics for the B-MMNL model. They show overall similar heterogeneity contributions of pasta compared to the MMNL and MNestL models, except for the Creamette brand. On the other hand, the heterogeneity contributions of sauce brand are generally of a smaller magnitude, suggesting a more homogeneous preference over pasta sauces across stores. 

\begin{table}[htbp]
	\centering
	\caption{B-MMNL Model Complementary Effects Between Pasta and Pasta Sauce}
	\label{tab:complementary_effects}
	\footnotesize
	\begin{tabular}{lccccc}
		\midrule\midrule
		                                  & \multicolumn{5}{c}{\textbf{Pasta sauce brands}} \\ \cline{2-6}
		                                  &         &         &               &             &\\
		                                  &  Ragu   &  Prego  & Private Label &    Hunt`s    & Mixed\\
		                                  &         &         &               &             &\\
		\textbf{\underline{Thin Spaghetti}} &         &         &               &             &\\
		Prv-Lab                    & -2.1477 &-2.1431 &-1.7780  &-1.8313 &-2.7933\\
		Barilla                        & -2.5435 &-2.3330&  -3.2412& -2.5694 &-2.3077  \\
		Mueller                     &-1.9418& -1.8820  &-3.2062 &-2.2658 &-2.7493   \\ 
		Creamette 				&-2.0614 &-1.9068 &-2.9909 &-2.3104 &-2.6289\\\bottomrule\bottomrule
	\end{tabular}
	 			\par\vspace{1ex}
	%	\raggedright
	\begin{minipage}{0.7\linewidth}
		\centering
		\footnotesize
		\begin{itemize}[]
\item Estimates are given for Thin Spaghetti, which is the most popular pasta type. The entries are the estimated $\gamma_r$ for bundle $r$, consisting of the pasta brand in the row and sauce in the column. 
		\end{itemize}
	\end{minipage}
\end{table}

Table~\ref{tab:complementary_effects} reports complementary effects for bundles including the Thin Spaghetti pasta, measured by the CVI variational mean of $\gamma_{r}$; See Part~E of the Online Appendix for the table of all complementary effects. 
Although one might anticipate some $\gamma_{r}>0$ if pasta and pasta sauce were routinely purchased together within a trip, storable categories typically exhibit intertemporal substitution shaped by household inventory dynamics \citep{hendelSalesConsumerInventory2006}. 
In this environment, $\gamma_{r}$ identifies contemporaneous co-purchase propensity, conditional on prices and controls, rather than complementarity \textit{per se}. 
Because households restock asynchronously---pasta and sauce inventories decrease at different rates---same-trip bundle purchases are relatively infrequent. 
Consistent with this mechanism, the negative $\gamma_{r}$ estimates mirror the high share of single-item transactions (over 70\%; see Table~\ref{tab:bundle_summary} in the Online Appendix), indicating that inventory-driven staggering dampens same-visit co-purchases even for economically complementary goods.

\section{Discussion}\label{sec: disc}
Currently available VI methods for estimating large mixed multinomial logit models of consumer choice have limitations. Data augmentation VI imposes strong independence assumptions between global parameters and random effects that can reduce estimation accuracy, while amortized VI can struggle to capture heterogeneity when group sizes are highly imbalanced, as a shared inference network must accommodate groups with very different information content.

The main contribution in this paper is the introduction of conjugating variational inference (CVI), a method that overcomes the limitations of existing approaches.
As in~\cite{tanUseModelReparametrization2021}, our new
method uses a 
second-order Taylor expansion to construct a variational approximation that accurately captures
dependence between the random coefficients and the model parameters. 
However, instead of solving of multiple inner optimization problems at each VI iteration, we introduce proxy auxiliary parameters that are refreshed only intermittently, yielding an approximation for the posterior of the random coefficients that is faster to compute. This allows the method to scale while remaining accurate.
We evaluate the method in simulations across standard, bundle, and nested mixed multinomial logit models in small and large samples. CVI outperforms competing methods in predictive accuracy for the first two and performs comparably for nested models, while also being significantly faster in the large sample illustrations.

We demonstrate the practical value of CVI in a large application to a pasta consumer choice dataset with over one million observations. The results show that: First, incorporating mixed effects in the standard and nested logit models substantially improves predictions of pasta brand choices. Second, store-level heterogeneity is driven primarily by heterogeneous preferences across brands rather than product type. Third, a profiling analysis shows that store-level own-price elasticities are negatively associated with store size and positively associated with average store-level transaction prices.

Another key contribution is the mixed multinomial bundle choice analysis in the pasta application. By augmenting pasta purchases with sauce choices, we show that the mixed bundle model improves predictive performance for pasta purchases, outperforming both the mixed multinomial and nested logit models. This is particularly relevant because mixed bundle models become computationally challenging as the number of choices grows due to the rapid increase in the dimension of the random effects and their covariance structure. 

Overall, the proposed CVI approach enables accurate and scalable estimation of mixed logit models in settings previously considered computationally challenging, including bundle choice models with high-dimensional random effects. Future research avenues include applying the methods to richer data environments, such as online retail consumer choice settings, and extending the flexibility of the approximation by allowing for skewness in the variational distribution of the random effects.

%\input{sec7}
%%\input{acknowl}
%\newpage
%\input{append}
\singlespacing
\newpage
%\bibliographystyle{apalike}
%\bibliography{references}
\printbibliography
%\input{tabs}
%\newpage
%\input{figs}
\FloatBarrier

\newpage
\onehalfspacing
\newpage
\noindent
\setcounter{page}{1}
\begin{center}
	{\bf \Large{Online Appendix for ``Conjugating Variational Inference for mixed multinomial choice models''}}
\end{center}

\vspace{10pt}

\setcounter{section}{0}
\setcounter{algorithm}{1}
\renewcommand{\thealgorithm}{\Alph{section}\arabic{algorithm}}
\noindent
This Online Appendix has five parts:

\begin{itemize}
	\item[] {\bf Part~A}: Notational conventions and matrix differentiation rules used.
	\item[] {\bf Part~B}: Additional details on the algorithms.
	\item[] {\bf Part~C}: Additional details on the simulation study.
	\item[] {\bf Part~D}: Additional details on the Carbo-Loading data.
	\item[] {\bf Part~E}: Additional details on the consumer choice application.
\end{itemize}
\newpage

\setcounter{figure}{0}
\setcounter{table}{0}
\renewcommand{\thetable}{A\arabic{table}}
\renewcommand{\thefigure}{A\arabic{figure}}

\noindent {\bf \Large{Part~A: Notational conventions and matrix differentiation rules used}}\\
\ \\
\noindent 
We outline the notational conventions that we adopt in computing
derivatives throughout the paper, which are the same as adopted in~\cite{loaiza-mayaFastAccurateVariational2022}. For a $d$-dimensional vector valued function $g(\bm x)$ of an $n$-dimensional
argument $\bm x$, $\frac{\partial g}{\partial \bm x}$ is the $d\times n$ matrix with element $(i,j)$ $\frac{\partial g_i}{\partial x_j}$.  This means for a scalar $g(\bm x)$, $\frac{\partial g}{\partial \bm x}$ is
a row vector.  When discussing the SGA algorithm we also sometimes write $\nabla_x g(\bm x)=\frac{\partial g}{\partial \bm x}^\top$, which is a column vector.
When the function $g(\bm x)$ or the argument $\bm x$ are matrix valued, then $\frac{\partial g}{\partial \bm x}$ is taken to 
mean $\frac{\partial \text{vec}(g(\bm x))}{\partial \text{vec}(\bm x)}$, where $\text{vec}(A)$ denotes the vectorization of a matrix $A$ obtained by stacking its columns one
underneath another.  If $g(x)$ and $h(x)$ are matrix valued functions, say $g(x)$ takes values which are $d\times r$ and $h(x)$ takes values which are $r\times n$, 
then a matrix valued product rule is
\begin{align*}
	\frac{\partial g(x)h(x)}{\partial x} & = (h(x)^\top\otimes I_d)\frac{\partial g(x)}{\partial x}+(I_n\otimes g(x))\frac{ \partial h(x)}{\partial x}
\end{align*}
where $\otimes$ denotes the Kronecker product and $I_a$ denotes the $a\times a$ identity matrix for a positive integer $a$.  

Some other useful results used repeatedly throughout the derivations below are
$$\text{vec}(ABC)=(C^\top\otimes A)\text{vec}(B),$$
for conformable matrices $A$, $B$ and $C$
the derivative 
\begin{align*}
	\frac{\partial A^{-1}}{\partial A} & = -(A^{-\top}\otimes A^{-1}).
\end{align*}
We also write $K_{m,n}$ for the commutation matrix (see, for example, Magnus and Neudecker, 1999).

Last, for scalar function $g(x)$ of scalar-valued argument $x$, we 
sometimes write $g^{\prime}(x)=\frac{d}{d x}g(x)$ and $g^{\prime\prime}(x)
=\frac{d^2}{d x^2}g(x)$ for the first and second derivatives with respect
to $x$ whenever it appears clearer to do so.
\newpage

\setcounter{figure}{0}
\setcounter{table}{0}
\renewcommand{\thetable}{B\arabic{table}}
\renewcommand{\thefigure}{B\arabic{figure}}

\noindent {\bf \Large{Part~B: Additional details on the algorithms}}\\
\ \\
\noindent  {\bf {B.1: DAVI algorithm}}\\
The following algorithm outlines the DAVI algorithm:
\begin{algorithm}[ht!]
	\begin{algorithmic}
		\State Initiate $\lambdavec^{(0)}$, $\lambdavec_i^{(0)}$ for $i = 1, \dots, S$. Generate $\thetavec^{(0)} \sim q_{\lambda^{(0)}}^0(\thetavec)$, $\alphavec^{(0)}_i \sim q_{\lambda i^{(0)}}(\alphavec_i)$ for $i = 1, \dots, S$. Set $t = 0$.
		\Repeat
		\State (a) Generate $\thetavec^{(t)} \sim q_{\lambda^{(t)}}^0(\thetavec)$  and  $\alphavec_i^{(t)} \sim q_{\lambda i^{(t)}
		}(\alphavec_i)$ for $i = 1, \dots, S$.
		\State (b) Compute the gradient with respect to $\lambdavec$ and $\lambdavec_i$ using reparametrization trick.
		\State (c) Compute step size $\boldsymbol{\rho}^{(t)}$ and $\boldsymbol{\rho}_i^{(t)}$  using an adaptive method (e.g. an ADA method)
		\State (d) Set $\lambdavec^{(t+1)} = \lambdavec^{(t)} + \boldsymbol{\rho}^{(t)} \circ \widetilde{\nabla}_{\lambda}\mathcal{L}(\lambdavec^{(t)})$ and $ \lambdavec_i^{(t+1)} = \lambdavec_i^{(t)} + \boldsymbol{\rho_i}^{(t)} \circ \widetilde{\nabla}_{\lambda i}\mathcal{L}(\lambdavec^{(t)})$ for $i = 1, \dots, S$.
		\State (e) Set $t = t+1$
		\Until{either a stopping rule is satisfied or a fixed number of steps is taken}
	\end{algorithmic}
	\caption{Data augmented variational inference}
	\label{alg:davi}
\end{algorithm}

\noindent  {\bf {B.2: Stopping rule for CVI and DAVI}}\\
The following algorithm outlines the stopping rule for CVI and DAVI algorithms:
\begin{algorithm}[ht!]
	\begin{algorithmic}
		\State Initialize \textit{Best\_ELBO} = -inf and \textit{count} = 0
		\If{$\text{mod}(t,100) = 0$ \textbf{and} $t > 1000$}
		\State Compute $\overline{ELBO} = \frac{1}{100} \sum\limits_{i=0}^{99} ELBO_{t-1000+10i}$
		\If{$\overline{ELBO} > \textit{Best\_ELBO}$}
		\State $\textit{Best\_ELBO} = \overline{ELBO}$
		\Else
		\State $\textit{count} = \textit{count} + 1$
		\EndIf
		\If{$\textit{count} > \textit{threshold}$}
		\State Stop the algorithm.
		\EndIf
		\EndIf
	\end{algorithmic}	\caption{Stopping rule for CVI and DAVI}
	\label{alg:stop_rule}
\end{algorithm}

\noindent  {\bf {B.3: Approximation to the conditional posterior}}\\
Recall the second order Taylor expansion:
\[
\log \tilde{p}(\bm{y}_i | \bm{\vartheta},\bm{\alpha}_i)
=
\log p(\bm{y}_i | \bm{\vartheta},\bm{a}_i)
+ \bm{g}_i^\top(\bm{\alpha}_i-\bm{a}_i)
- \tfrac{1}{2}(\bm{\alpha}_i-\bm{a}_i)^\top H_i (\bm{\alpha}_i-\bm{a}_i),
\]
where $\bm{g}_i$ and $H_i$ are the gradient and (negative) Hessian of 
$\log p(\bm{y}_i | \bm{\vartheta},\bm{\alpha}_i)$ evaluated at $\bm{\alpha}_i=\bm{a}_i$.
This gives
\begin{align*}
	\tilde{p}(\bm{y}_i | \bm{\vartheta},\bm{\alpha}_i)& = p(\bm{y}_i | \bm{\vartheta},\bm{a}_i)\exp\left(-\frac{1}{2}\left[\alphavec_i^\top H_i\alphavec_i - 2 \alphavec_i^\top\vvec_i\right] - \frac{1}{2}\avec_i^\top H_i\avec_i \right)
\end{align*}
Here $\vvec_i = \gvec_i + H_i \avec_i$. If $\tilde{p}(\bm{y}_i | \bm{\vartheta},\bm{a}_i)$ consists of multiple observations, then $ \gvec_i =  \sum_{t = 1}^{T_i}\gvec_{it}$ and $H_i =  \sum_{t = 1}^{T_i}H_{it}$ where $T_i$ is the number of observations in group $i$. The approximated conditional posterior for $\alphavec_i$ can be constructed as:
\begin{align*}
	q(\alphavec_i\mid \varthetavec,\avec_i) &\propto\tilde{p}(\bm{y}_i | \bm{\vartheta},\bm{\alpha}_i) p(\alphavec_i;\varthetavec)
\end{align*}
where $ p(\alphavec_i;\varthetavec) = \phi(\alphavec_i; \xivec,\Sigma)$ and $q(\alphavec_i\mid \varthetavec,\avec_i) = \phi(\alphavec_i; \muvec_i, V_i)$ with $V_i = (H_i + \Sigma^{-1})^{-1}$ and $\muvec_i = V_i (\vvec_i+ \Sigma^{-1}\xivec)$. Note that both $\xivec$ and $\Sigma$ in $q(\alphavec_i\mid \varthetavec,\avec_i)$ are constructed from $\varthetavec$, not the variational parameters $\thetavec$. \\

\noindent  {\bf {B.4: Derivation of re-parametrized ELBO gradient}}\\
Consider the generative formula $\bm{\theta} = h(\bm{\epsilon},\bm{\lambda})$ for $q_{\lambda_0}(\bm{\theta})$, where $\bm{\epsilon}\sim f_{\epsilon}(\bm{\epsilon})$ and $f_{\epsilon}(\bm{\epsilon})$ is a distribution that does not depend on $\bm{\lambda}$. 
With this change of variable, the ELBO expression 
\begin{equation*}
	\mathcal{L}\left(\bm{\lambda}\right) = E_{q_{\lambda_0}({\theta})q(\bm{\alpha}|\bm{\vartheta},\bm{a})}\left[\log p(\bm{y}|\bm{\psi})p(\bm{\psi})-\log q_\lambda(\bm{\theta})q(\bm{\alpha}|\bm{\vartheta},\bm{a}))\right],
\end{equation*}
can be alternatively written as
\begin{equation*}
	\mathcal{L}\left(\bm{\lambda}\right) = E_{f_{\epsilon}(\epsilon)q(\bm{\alpha}|\bm{\vartheta},\bm{a})}\left[\log p(\bm{y}|\bm{\alpha},h(\bm{\epsilon},\bm{\lambda}))p(h(\bm{\epsilon},\bm{\lambda}),\bm{\alpha})-\log q_\lambda(h(\bm{\epsilon},\bm{\lambda}))q(\bm{\alpha}|\bm{\vartheta},\bm{a}))\right].
\end{equation*}
The derivative of this expression with respect to $\bm{\lambda}$ is
\begin{align*}
	\frac{\partial}{\partial\bm{\lambda}}\mathcal{L}\left(\bm{\lambda}\right) &= E_{f_{\epsilon}(\epsilon)q(\bm{\alpha}|\bm{\vartheta},\bm{a})}\left[\frac{\partial}{\partial\bm{\theta}} \left[\log p(\mathbf{y}|\bm{\psi})p(\bm{\psi})\right]\frac{\partial\bm{\theta}}{\partial\bm{\lambda}}-\frac{\partial}{\partial\bm{\theta}}[\log q_\lambda(\bm{\theta})q(\bm{\alpha}|\bm{\vartheta},\bm{a})]\frac{\partial\bm{\theta}}{\partial\bm{\lambda}}\right]\\
	&= E_{f_{\epsilon}(\epsilon)q(\bm{\alpha}|\bm{\vartheta},\bm{a})}\left[ \left\{\frac{\partial}{\partial\bm{\theta}}\left[\log p(\mathbf{y}|\bm{\psi})p(\bm{\psi})\right]-\frac{\partial}{\partial\bm{\theta}}[\log q_\lambda(\bm{\theta})q(\bm{\alpha}|\bm{\vartheta},\bm{a})]\right\}\frac{\partial\bm{\theta}}{\partial\bm{\lambda}}\right]\\
	&= E_{f_{\epsilon}(\epsilon)q(\bm{\alpha}|\bm{\vartheta},\bm{a})}\left[ \left\{\frac{\partial}{\partial\bm{\theta}}\left[\log p(\mathbf{y}|\bm{\psi})p(\bm{\psi})\right]-\frac{\partial}{\partial\bm{\theta}}[\log q_\lambda(\bm{\theta})]\right\}\frac{\partial\bm{\theta}}{\partial\bm{\lambda}}\right].
\end{align*}
To derive the gradient of the ELBO we must simply transpose the expression above to get:
\begin{align*}
	\nabla_\lambda\mathcal{L}\left(\bm{\lambda}\right) &= E_{f_{\epsilon}(\epsilon)q(\bm{\alpha}|\bm{\vartheta},\bm{a})}\left[\frac{\partial\bm{\theta}}{\partial\bm{\lambda}}^\top\left\{\frac{\partial}{\partial\bm{\theta}}\left[\log p(\mathbf{y}|\bm{\psi})p(\bm{\psi})\right]-\frac{\partial}{\partial\bm{\theta}}[\log q_\lambda(\bm{\theta})]\right\}^\top\right]\\
	& = E_{f_{\epsilon}(\epsilon)q(\bm{\alpha}|\bm{\vartheta},\bm{a})}\left[\frac{\partial\bm{\theta}}{\partial\bm{\lambda}}^\top\left\{\nabla_\theta\log p(\mathbf{y}|\bm{\psi})p(\bm{\psi})-\nabla_\theta\log q_\lambda(\bm{\theta})\right\}\right].
\end{align*}

\noindent{\bf{B.5: inference details of mixed multinomial logit model}}\\
The log-likelihood function for observation $i$ at time $t$ with $Y_{it} = j$ is
\begin{align*}
	\ell_{it} = v_{itj} - \log \sum_{j' = 1}^{J}\exp(v_{itj'})
\end{align*}
The gradient of $\ell_{it}$ with respect to $\alphavec_{ij}$ is
\begin{align*}
	\frac{\partial \ell_{it}}{\partial \alphavec_{ij}} = (y_{itj} - p_{itj})\xvec_{itj}
\end{align*}
The second order derivative of $\ell_{it}$ w.r.t $\alphavec_{ij}$ and $\alphavec_{il}$ has two scenarios:
\begin{align}
	\frac{\partial^2 \ell_{it}}{\partial \alphavec_{ij}\partial \alphavec_{il}} & = \begin{cases}
		& (p_{itj}\xvec_{itj})(p_{itl}\xvec_{itl})^\top \quad \text{ if } l\neq j\\
		& (p_{itj}\xvec_{itj})(p_{itl}\xvec_{itl})^\top - p_{itj} \xvec_{itj}\xvec_{itj}^\top \quad \text{ if } l = j
	\end{cases}
\end{align}
\noindent{\bf{B.6: inference details of mixed nested logit model}}\\

The log-likelihood function for observation $i$ at time $t$ with $Y_{it} = j$ and $j \in B_k$ is:
\begin{align}
	\ell_{it} &= v_{itj}/\tau_k + (\tau_k -1)\log \left(\sum_{j'\in B_k} \exp(v_{itj'/\tau_k})\right) - \log\left(\sum_{l = 1}^K\left(\sum_{j''\in B_l}(\exp(V_{itj''}/\tau_l))\right)^{\tau_l}\right) 
\end{align} 
The gradient of $\ell_{it}$ w.r.t $\alphavec_{ir}$ has three scenarios:
\begin{align}
	\frac{\partial \ell_{it}}{\partial \alphavec_{ir}} & = \begin{cases}
		&\xvec_{itj}/\tau_k + (\tau_k -1)\frac{\exp(v_{itj}/\tau_k)\xvec_{itj}/\tau_k}{\sum_{j'\in B_k} \exp(v_{itj'/\tau_k})} - \frac{\tau_k(\sum_{j'\in B_k}(\exp(V_{itj'}/\tau_k)))^{\tau_k -1}\exp(V_{itj}/\tau_k)(\xvec_{itj}/\tau_k)}{\sum_{l = 1}^K\left(\sum_{j''\in B_l}(\exp(V_{itj''}/\tau_l))\right)^{\tau_l}} \quad \text{ if } r =j \nonumber \\
		&(\tau_k -1)\frac{\exp(v_{itr}/\tau_k)\xvec_{itr}/\tau_k}{\sum_{j'\in B_k} \exp(v_{itj'/\tau_k})} - \frac{\tau_k(\sum_{j'\in B_k}(\exp(V_{itj'}/\tau_k)))^{\tau_k -1}\exp(V_{itr}/\tau_k)(\xvec_{itr}/\tau_k)}{\sum_{l = 1}^K\left(\sum_{j''\in B_l}(\exp(V_{itj''}/\tau_l))\right)^{\tau_l}} \quad \text{ if } r \neq j \text{ and } r\in B_k \nonumber \\
		& - \frac{\tau_m(\sum_{j'\in B_m}(\exp(V_{itj'}/\tau_m)))^{\tau_m -1}\exp(V_{itr}/\tau_m)(\xvec_{itr}/\tau_m)}{\sum_{l = 1}^K\left(\sum_{j''\in B_l}(\exp(V_{itj''}/\tau_l))\right)^{\tau_l}} \quad \text{ if } r\notin B_k \\
	\end{cases}
\end{align}
Similarly, the block diagonal elements of Hessian of $\ell_{it}$ w.r.t $\alphavec_{ir}$ has three scenarios. For $r \in B_k$:
{\footnotesize
	\begin{align}
		\frac{\partial^2 \ell_{it}}{\partial \alphavec_{ir}\partial \alphavec_{ir}^\top} & = 
		(\tau_k -1)\left[-\frac{(\exp(v_{itr}/\tau_k))(\xvec_{itr}/\tau_k)(\xvec_{itr}/\tau_k)^\top(\exp(v_{itr}/\tau_k))}{(\sum_{j'\in B_k} \exp(v_{itj'/\tau_k}))^2}  + \frac{\exp(v_{itr}/\tau_k)(\xvec_{itr}/\tau_k)(\xvec_{itr}/\tau_k)^\top}{\sum_{j'\in B_k} \exp(v_{itj'/\tau_k})} \right] \nonumber\\
		& \quad + \frac{(\tau_k(\sum_{j'\in B_k}(\exp(V_{itj'}/\tau_k)))^{\tau_k -1}\exp(V_{itr}/\tau_k))(\xvec_{itr}/\tau_k)(\xvec_{itr}/\tau_k)^\top(\tau_k(\sum_{j'\in B_k}(\exp(V_{itj'}/\tau_k)))^{\tau_k -1}\exp(V_{itr}/\tau_k)) }{\left(\sum_{l = 1}^K\left(\sum_{j''\in B_l}(\exp(V_{itj''}/\tau_l))\right)^{\tau_l}\right)^2}  \nonumber\\
		& \quad - \frac{\tau_k(\tau_k-1)(\sum_{j'\in B_k}(\exp(V_{itj'}/\tau_k)))^{\tau_k -2}\exp(V_{itr}/\tau_k)(\xvec_{itr}/\tau_k)(\xvec_{itr}/\tau_k)^\top\exp(V_{itr}/\tau_k)}{\sum_{l = 1}^K\left(\sum_{j''\in B_l}(\exp(V_{itj''}/\tau_l))\right)^{\tau_l}}  \text{ Same nest}\nonumber\\
		& \quad - \frac{\tau_k(\sum_{j'\in B_k}(\exp(V_{itj'}/\tau_k)))^{\tau_k -1}\exp(V_{itr}/\tau_k)(\xvec_{itr}/\tau_k)(\xvec_{itr}/\tau_k)^\top}{\sum_{l = 1}^K\left(\sum_{j''\in B_l}(\exp(V_{itj''}/\tau_l))\right)^{\tau_l}}  \text{ diagonal only}
	\end{align}
}
For $r\notin B_k$, the Hessian only includes the last three lines. For numerical stability of the calculation, we use log-sum-exp trick where possible. For $\sum_{l = 1}^K\left(\sum_{j''\in B_l}(\exp(V_{itj''}/\tau_l))\right)^{\tau_l}$ we subtract the maximum $c = \text{argmax} V_{itj}$ such that $\sum_{l = 1}^K\left(\sum_{j''\in B_l}(\exp(V_{itj''}/\tau_l))\right)^{\tau_l} = c \tau_{max} + \log \left(\sum_{l = 1}^{K}\exp(c(\tau_l - \tau_{max})(\sum_{j''\in B_l} (\exp (V_{itj}/\tau_l-c))^{\tau_l}))\right)$.

\clearpage
\FloatBarrier

% PART C
\setcounter{figure}{0}
\setcounter{table}{0}
\renewcommand{\thetable}{C\arabic{table}}
\renewcommand{\thefigure}{C\arabic{figure}}

\noindent {\bf \Large{Part~C: Additional details on the simulation study.}}\\
\ \\
\noindent  {\bf {C.1: Data generating process of the mixed multinomial logit model}}\\
\begin{algorithm}
	\caption{Data generating process}\label{alg:dgp}
	\begin{algorithmic}
		\For{$i = 1, \dots, S$}
		\State (a) Draw $\alphavec_i \sim N(\xivec,\Sigma)$.
		\For{$t = 1,\dots, T_i$}
		\State (b) Denote  $X_{it} = (\mathbf{1}_{J-1}, \tilde{X}_{it})^\top$, $\tilde{X}_{it} =(\tilde{\xvec}_{it2},\dots,\tilde{\xvec}_{itJ})^\top$, $\tilde{\xvec}_{itj} \sim %N(\mathbf{0},\mI_{m-1})$ for $j = 2,\dots,J$.
		U_{m-1}(0,1)$ for $j = 2,\dots,J$.
		\State (c) Denote  $\uvec_{it} = (0, (\betavec^\top ,\alphavec_i^\top) X_{it}^\top)^\top$ and $\pvec_{itj} = \frac{\exp(u_{itj})}{\sum_{j' = 1}^{J}\exp(u_{itj'})}$.
		\State (d) Draw $y_{it} \sim \text{Cat}(J,\pvec_{it})$.
		\EndFor
		\EndFor
	\end{algorithmic}
	\label{dgp:mmnl}
\end{algorithm}
Here $J$ is the number of alternatives, $m$ is the number of covariates including the intercept term. The values of the true parameters are
\begin{align*}
	\betavec &= \begin{bmatrix}
		0  &  0  &  0  \\
		-0.10 & 0.35& -0.15\\
		-0.15& 0.3 &  0.40 \\
		0.40 & -0.15& 0.58
	\end{bmatrix}\\
	\xivec& = \begin{bmatrix}
		-0.9640&0.4002& -0.3788&0.2409
	\end{bmatrix}^\top
\end{align*}
We generate the covariance matrix $\Sigma$ by drawing a Cholesky factor $L_{\text{corr}}$ from an LKJ distribution with shape parameter $1$, implying a uniform distribution over correlation matrices. We set $\Sigma = \operatorname{diag}(\mathbf{1}) \, R \, \operatorname{diag}(\mathbf{1})$, resulting a positive definite covariance matrix with unit marginal variances. For each model, $\Sigma$ is held fixed across replications so that all datasets are generated from the same DGP. 
\newpage

\noindent  {\bf {C.2: Computing the predictive metrics}}\\
In this paper, we report the log-score and the weighted macro F1 score ($\mathbb{F}1$) as measures of predictive accuracy. Let $p_{itj}$ denote the predictive probability of alternative $j$ being chosen by decision maker $i$ at time $t$. The log-score (LS) is defined as:
\begin{align}
	LS = \frac{ \sum_{i = 1}^{S} \sum_{t = 1}^{T_i} \log\left( \yvec_{it}^\top \pvec_{it} \right) }{ \sum_{i = 1}^{S} T_i }
	\label{eq:ls}
\end{align}
Here, $\pvec_{it} = (p_{it1}, \dots, p_{itJ})^\top$ is the vector of predictive choice probabilities over the $J$ alternatives, and $\yvec_{it} = (y_{it1}, \dots, y_{itJ})^\top$ is the one-hot encoded true outcome vector, with $y_{itj} = 1$ if $Y_{it} = j$, and $y_{itj’} = 0$ for all $j’ \ne j$.

The predicted class label for each observation is given by:
\begin{align}
	Y^{\text{pred}}_{it} = \arg\max_j \left( \pvec_{it} \right)
	\label{eq:ypred}
\end{align}
The $\mathbb{F}1$ score is computed using the f1\_score function from the scikit-learn \citep{scikit-learn} package, with the predicted labels $Y^{\text{pred}}_{it}$ and the true labels $Y_{it}$ as inputs.

Since the predictive choice probability 
\[ p_{itj} = \int_{\thetavec, \alphavec} p_{itj}(\thetavec, \alphavec)\, q_{\lambdavec^*}(\thetavec, \alphavec) \, d\thetavec \, d\alphavec
\] 

is analytically intractable, we approximate $p_{itj}$ using Monte Carlo simulation:
\begin{align}
	\hat{p}_{itj} = \frac{1}{N_{\text{sim}}} \sum_{\text{iter}=1}^{N_{\text{sim}}} p_{itj}(\thetavec^{(\text{iter})}, \alphavec^{(\text{iter})}), \quad \text{where } (\thetavec^{(\text{iter})}, \alphavec^{(\text{iter})}) \sim q_{\lambdavec^*}(\thetavec, \alphavec)
\end{align}
Here, $q_{\lambdavec^*}(\thetavec, \alphavec)$ is the calibrated variational posterior. The approximated probabilities $\hat{p}_{itj}$ are then used in equations (\ref{eq:ls}) and (\ref{eq:ypred}) to compute the log-score and $\mathbb{F}1$ score, respectively.

Similarly, we can approximate ELBO using Monte Carlo simulation:
\begin{align}
	\widehat{\text{ELBO}} = \frac{1}{N_{\text{sim}}} \sum_{\text{iter}=1}^{N_{\text{sim}}} \left[ \log p(\yvec \mid \thetavec^{(\text{iter})}, \alphavec^{(\text{iter})}) + \log p(\thetavec^{(\text{iter})}, \alphavec^{(\text{iter})}) - \log q_{\lambdavec^*}(\thetavec^{(\text{iter})}, \alphavec^{(\text{iter})}) \right]
\end{align}

\noindent  {\bf {C.3: Additional results from the simulation studies}}\\

\begin{figure}[htbp]
	\centering
	\begin{subfigure}[b]{0.4\textwidth}
		\centering
		\includegraphics[width=0.9\textwidth]{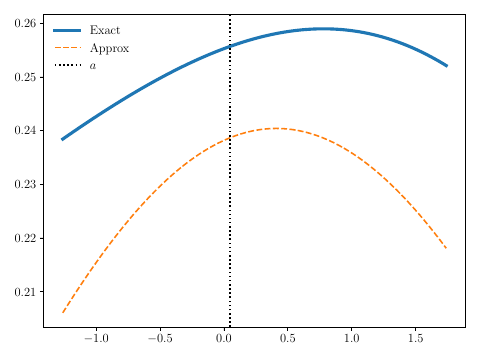}
		\caption{ $iter = 0$}
	\end{subfigure}
	\hfill
	\begin{subfigure}[b]{0.4\textwidth}
		\centering
		\includegraphics[width=0.9\textwidth]{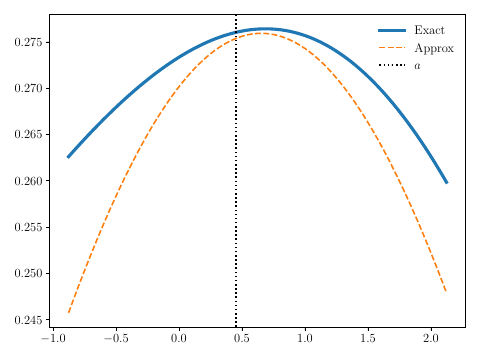}
		\caption{ $iter = 350$}
	\end{subfigure}
	\hfill
	\begin{subfigure}[b]{0.4\textwidth}
		\centering
		\includegraphics[width=0.9\textwidth]{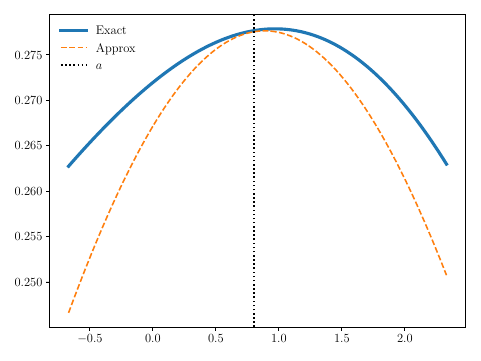}
		\caption{ $iter = 1000$}
	\end{subfigure}
	\hfill
	\begin{subfigure}[b]{0.4\textwidth}
		\centering
		\includegraphics[width=0.9\textwidth]{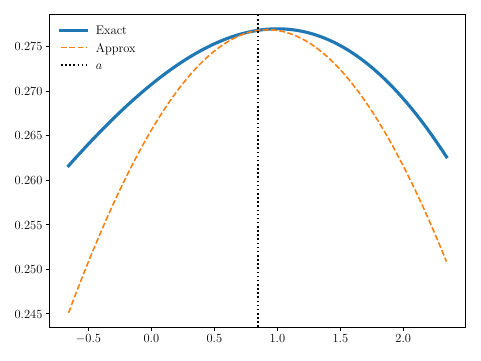}
		\caption{ $iter = 2000$}
	\end{subfigure}
	\caption{Approximation accuracy of the second-order Taylor expansion for a representative group from Simulation~1 at different optimization steps $iter$. The exact likelihood is evaluated at $\alphavec^{(iter)}$, while the approximated likelihood is constructed using a second-order Taylor expansion around $\avec^{(iter)}$ and also evaluated at $\alphavec^{(iter)}$. The value of $\alpha^{(iter)}_{(3,2)}$ is varied along the grid to trace out the likelihood profiles, while all other coefficients remain fixed.}
	\label{fig:likelihood_approx_mmnl}
\end{figure}

\begin{table}[thb!]
	\captionsetup{skip=0pt}
	\centering
	\caption{Predictive accuracy for the MMNL model using simulated data}
	\footnotesize
	\begin{tabular}{ccccc}
		\toprule
		Method & \makecell{Log-score\\(train)} & \makecell{$\mathbb{F}1$\\(train)}& \makecell{Log-score\\(test)} & \makecell{$\mathbb{F}1$\\(test)}\\
		\multicolumn{2}{l}{	E.g. 2(a): smaller example}&&&\\
		\cmidrule(r){1-2}
		{CVI}  &-1.1563 &0.4719 &{-1.1860} &{0.4587} \\
		{CVI-lkj}  & {-1.1528} &{0.4721} &-1.1867 &{0.4566} \\
		{AVI-lkj}   & {-1.1551} &{0.4728}&-1.1875 &0.4567 \\
		{DAVI} &-1.1708 &0.4647&-1.1867 &0.4610\\
		{Na\"ive} & -1.3700 &0.1628&-1.3712 &0.1577\\
		
		\multicolumn{2}{l}{	E.g. 2(b): larger example}&&&\\
		\cmidrule(r){1-2}
		{CVI} & {-1.1453} &{0.4872}&{-1.1880} &{0.4611}\\
		{CVI-lkj}  & -1.1452 &0.4872&-1.1880 &0.4610\\
		{AVI-lkj} & -1.1551 &0.4809&-1.1904 &0.4595  \\
		{DAVI} &-1.1582 &0.4793 &-1.1930 &0.4580\\
		{Na\"ive} & -1.3712 &0.1530&-1.3713 &0.1530 \\
		\bottomrule
	\end{tabular}
	\par\vspace{1ex}
	%	\raggedright
	\begin{minipage}{0.55\linewidth}
		\centering
		\footnotesize
		\begin{itemize}[]
			\item Higher values of both $\mathbb{F}1$ score and Log-score mean better predictive accuracy.
			\item $\mathbb{F}1$ scores and Log-scores are evaluated based on $3000$ simulations for the smaller example and $1000$ simulations for the larger example.
		\end{itemize}
	\end{minipage}
	\label{tab: results_MMNL_predict}
\end{table}

\begin{table}[thb!]
	\captionsetup{skip=0pt}
	\centering
	\caption{Predictive accuracy for the MNestL model using simulated data}
	\footnotesize
	\begin{tabular}{ccccc}
		\toprule
		Method & \makecell{Log-score\\(train)} & \makecell{$\mathbb{F}1$\\(train)}& \makecell{Log-score\\(test)} & \makecell{$\mathbb{F}1$\\(test)}\\
		\multicolumn{2}{l}{	E.g. 3(a): smaller example}&&&\\
		\cmidrule(r){1-2}
		{CVI}  &-0.7941 &0.6454&-0.8210 &0.6459 \\
		{CVI-lkj}  & -0.7898 &0.6483&-0.8177&0.6464\\
		{AVI-lkj}   &{-0.7862} &{0.6573}&{-0.8112} &0.6503\\
		{DAVI} &{-0.7937} &0.6491&-0.8277 &{0.6550}\\
		{Na\"ive} & -1.3360 &0.0523&-1.3293 &0.0452\\
		
		\multicolumn{2}{l}{	E.g. 3(b): larger example}&&&\\
		\cmidrule(r){1-2}
		{CVI} &-0.8237 &0.6323&-0.8352&0.6284\\
		{CVI-lkj}  &  -0.8238 &0.6323&-0.8352&0.6283\\
		{AVI-lkj} & -0.8313& 0.6291&-0.8401&0.6259\\
		{DAVI} &{-0.8234}&{0.6331}&{-0.8345}&{0.6293}\\
		SVI-FE&-1.3193& 0.3008&-1.3192&0.3009\\
		{Na\"ive} & -1.3473&0.0568&-1.3476&0.0568\\
		\bottomrule
	\end{tabular}
	\par\vspace{1ex}
	\begin{minipage}{0.55\linewidth}
		\centering
		\footnotesize
		\begin{itemize}[]
			\item Higher values of both $\mathbb{F}1$ score and Log-score mean better predictive accuracy.
			\item $\mathbb{F}1$ scores and Log-scores are evaluated based on $3000$ simulations for the smaller example and $1000$ simulations for the larger example.
		\end{itemize}
	\end{minipage}	
	\label{tab: results_mixnl_predict}
\end{table}

\begin{table}[thb!]
	\captionsetup{skip=0pt}
	\centering
	\caption{Predictive accuracy for the B-MMNL model using simulated data}
	\footnotesize
	\begin{tabular}{ccccc}
		\toprule
		Method & \makecell{Log-score\\(train)} & \makecell{$\mathbb{F}1$\\(train)}& \makecell{Log-score\\(test)} & \makecell{$\mathbb{F}1$\\(test)}\\
		\multicolumn{2}{l}{	E.g. 2(a): smaller example}&&&\\
		\cmidrule(r){1-2}
		{CVI}  &-1.7511 &{0.2621}&{-1.7968} &0.2523 \\
		{CVI-lkj}  & {-1.7504} &{0.2609}  &{-1.7979} &0.2507\\
		{AVI-lkj}   &-1.7498 &0.2618&-1.7984 &{0.2527}\\
		{DAVI} &-1.7662 &0.2555 &-1.8016 &0.2538\\
		{Na\"ive} & -2.0225 &0.0629 &-2.0302 &0.0615\\
		
		\multicolumn{2}{l}{	E.g. 2(b): larger example}&&&\\
		\cmidrule(r){1-2}
		{CVI} & {-1.7964} &{0.2395}&{-1.8451} &{0.2228}\\
		{CVI-lkj}  &  {-1.7967} &{0.2391}&{-1.8451} &{0.2226}\\
		{AVI-lkj} & -1.8113 &0.2338&-1.8492 &0.2213  \\
		{DAVI} &-1.8044 &0.2361&-1.8480 &0.2216\\
		{Na\"ive} & -2.0266 &0.1886&-2.0275 &0.1879 \\
		\bottomrule
	\end{tabular}
	\par\vspace{1ex}
	%	\raggedright
	\begin{minipage}{0.55\linewidth}
		\centering
		\footnotesize
		\begin{itemize}[]
			\item Higher values of both $\mathbb{F}1$ score and Log-score mean better predictive accuracy.
			\item $\mathbb{F}1$ scores and Log-scores are evaluated based on $3000$ simulations for  the smaller example and $1000$ simulations for the larger example.
		\end{itemize}
	\end{minipage}
	\label{tab: results_mixbc_predict}
\end{table}

  \begin{figure}
	\centering
	\includegraphics[width=1\linewidth]{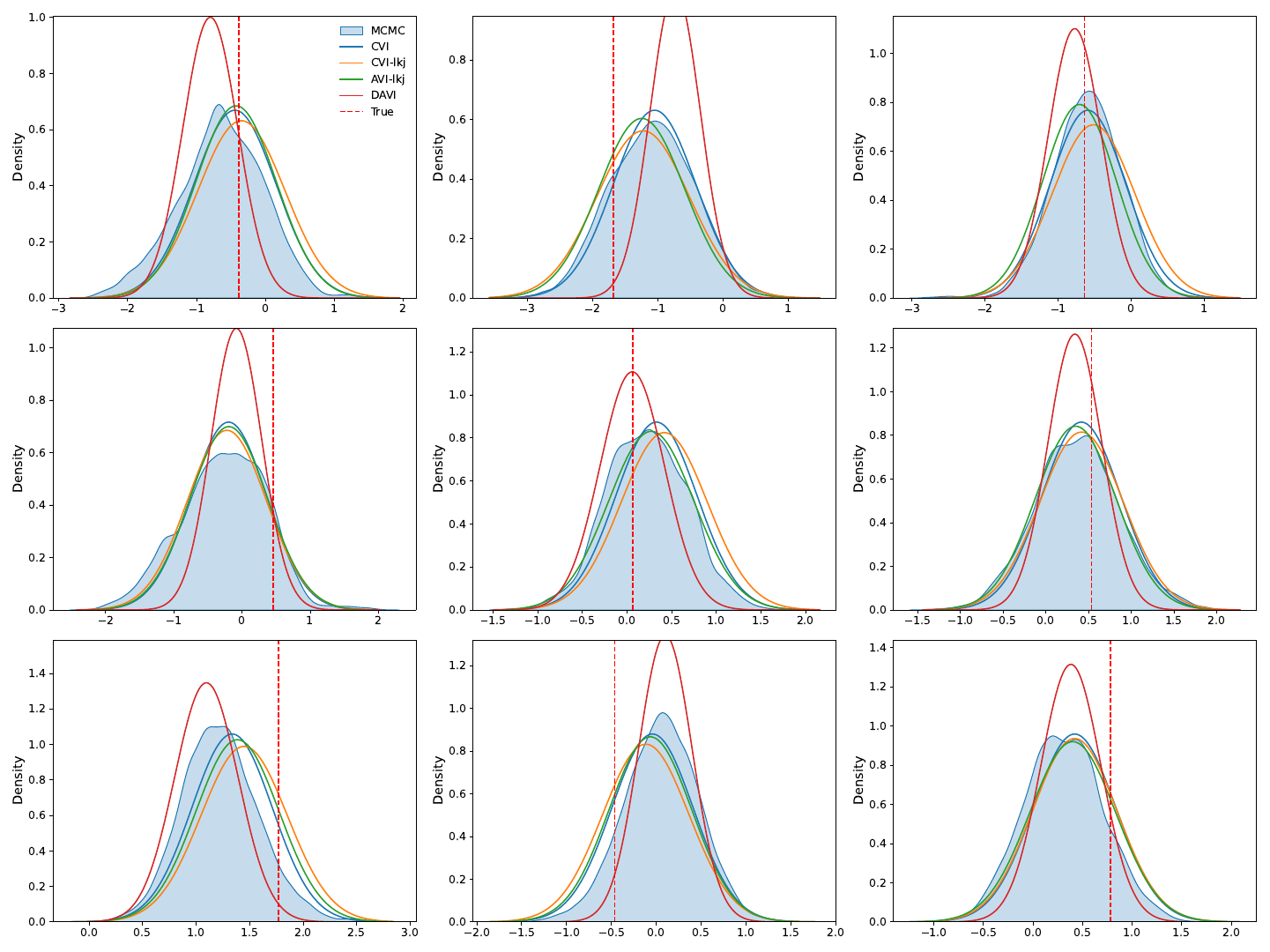}
	\caption{Posteriors of random coefficients of a representative group for the smaller simulated B-MMNL example: The figure displays the posterior distributions obtained using three VI methods and MCMC. Each row represents the random coefficients associated with a specific alternative, while each column corresponds to the random coefficients of a specific covariate across alternatives. For identification purposes, the coefficients of the first (reference) alternative are fixed at $0$.}
	\label{fig:posterior_alpha_mixbc}
\end{figure}
 \begin{table}[ht!]
	\begin{center}
		\caption{Estimates of Heterogeneity in Simulation~2}
		\label{tab:Mixbc_sim_TH}
		\begin{tabular}{cccccccccc}
			\hline
			&
			& \multicolumn{4}{c}{E.g.~2(a): Smaller Example} 
			& \multicolumn{4}{c}{E.g.~2(b): Larger Example} \\
			\cmidrule(lr){3-6} \cmidrule(lr){7-10}
			Method 
        & $p(\Sigma)$ 
		& TH & AH(2) & AH(3) & AH(4) 
		& TH & AH(2) & AH(3) & AH(4) \\
		\hline
		True    
		& {} 
		& 23.520 & 1.412 & 1.540 & 1.544 
		& 11.424 & 1.088 & 1.212 & 1.080 \\
		{MCMC}  
		& {HW}
		& 22.512 & 1.372 & 1.380 & 1.612
		& -- & -- & -- & -- \\
		CVI     
		& {HW} 
		& 19.936 & 1.212 & 1.192 & 1.444 
		& 10.696 & 1.052 & 1.164 & 0.992 \\
		CVI
		& {LKJ} 
		& 23.016 & 1.448 & 1.360 & 1.568 
		& 10.584 & 1.048 & 1.148 & 0.988 \\
		AVI
		& {LKJ} 
		& 22.848 & 1.376 & 1.400 & 1.628 
		& 10.136 & 1.020 & 1.168 & 0.988 \\
		DAVI    
		& {HW} 
		& 18.144 & 1.104 & 1.060 & 1.372 
		& 9.968 & 1.060 & 1.140 & 0.984 \\
			\hline
		\end{tabular}
	\end{center}
	Note: Estimates are based on $3{,}000$ and $1{,}000$ simulated $\Sigma$ for the smaller and larger examples, respectively.
\end{table}

 \begin{table}[ht!]
	\begin{center}
		\caption{Estimates of Complementarity in Simulation~2}
		\label{tab:Mixbc_gamma_sim}
		\begin{tabular}{cccccccccc}
			\hline
			&
			& \multicolumn{4}{c}{E.g.~2(a): Smaller Example} 
			& \multicolumn{4}{c}{E.g.~2(b): Larger Example} \\
			\cmidrule(lr){3-6} \cmidrule(lr){7-10}
			Method 
			&  			 $p(\Sigma)$ 
			& $\gamma_5$ & $\gamma_6$ & $\gamma_7$ & $\gamma_8$
			& $\gamma_5$ & $\gamma_6$ & $\gamma_7$ & $\gamma_8$\\
			\hline
			True    
			& {} 
			& 0.098 & 0.430 & 0.206 & 0.090 
			& 0.098 & 0.430 & 0.206 & 0.090 \\
			{MCMC}  
			&{HW}
			& 0.180   &0.387  &0.220  &-0.014
			& -- & --& -- & --\\
			CVI     
			& {HW} 
			& 0.176  &0.404  &0.198 &-0.018
			& 0.103 &0.438 &0.172 &0.078 \\
			CVI
			& {LKJ} 
			& 0.180   &0.410   &0.194 &-0.017 
			& 0.103 &0.439 &0.175 &0.087 \\
			AVI
			& {LKJ} 
			& 0.195  & 0.393  & 0.236 & -0.015
			& 0.096 & 0.419 & 0.214 & 0.098 \\
			DAVI    
			& {HW} 
			& 0.181  &0.383  &0.229 &-0.001
			& 0.103 &0.425 &0.201 &0.096 \\
			\hline
		\end{tabular}
	\end{center}
	Note: MCMC estimates are sample means, VI estimates are variational mean of calibrated VAs.
\end{table}

\begin{table}[ht!]
	\begin{center}
		\caption{Estimates of Heterogeneity in Simulation~3}
		\label{tab:MixNL_sim_TH}
		\begin{tabular}{cccccccccc}
			\hline
			& 
			& \multicolumn{4}{c}{E.g.~3(a): Smaller Example} 
			& \multicolumn{4}{c}{E.g.~3(b): Larger Example} \\
			\cmidrule(lr){3-6} \cmidrule(lr){7-10}
			Method 
			& $p(\Sigma)$ 
			& TH & AH(2) & AH(3) & AH(4) 
			& TH & AH(2) & AH(3) & AH(4) \\
			\hline
			True    
			& {--} 
			& 36.232 & 2.432 & 2.608 & 2.728 
			& 28.448 & 2.456 & 2.808 & 2.240 \\
			CVI     
			& {HW} 
			& 14.280 & 0.440 & 0.880 & 1.244 
			& 20.272 & 1.448 & 2.096 & 1.612 \\
			CVI
			& {LKJ} 
			& 16.240 & 0.648 & 1.020 & 1.432 
			& 20.048 & 1.428 & 2.096 & 1.596 \\
			AVI     
			& {LKJ} 
			& 36.512 & 3.488 & 2.364 & 2.572 
			& 26.040 & 2.184 & 2.520 & 2.012 \\
			DAVI    
			& {HW} 
			& 20.720 & 1.948 & 1.548 & 2.004 
			& 28.448 & 2.480 & 2.876 & 2.204 \\
			\hline
		\end{tabular}
	\end{center}
	Note: Estimates are based on $3{,}000$ simulated $\Sigma$ for the smaller example and $1{,}000$ simulations for the larger example. We do not have MCMC results in the smaller example as MCMC failed to converge.
\end{table}

 \begin{table}[ht!]
	\begin{center}
		\caption{Estimates of Nesting Parameters in Simulation~3}
		\label{tab:MixNL_tau_sim}
		\begin{tabular}{cccccc}
			\hline
			&
			& \multicolumn{2}{c}{E.g.~2(a): Smaller Example} 
			& \multicolumn{2}{c}{E.g.~2(b): Larger Example} \\
			\cmidrule(lr){3-4} \cmidrule(lr){5-6}
			Method 
			&  			 $p(\Sigma)$ 
			& $\tau_1$ & $\tau_2$ & $\tau_1$ & $\tau_2$\\
			\hline
			True    
			& {} 
			& 0.300 & 0.700 & 0.300 & 0.700 \\
			CVI     
			& {HW} 
			& 0.124  &0.257  &0.234 &0.562 \\
			CVI
			& {LKJ} 
			& 0.140&  0.307   &0.235 &0.564  \\
			AVI
			& {LKJ} 
			& 0.425 &0.732  & 0.356 & 0.700 \\
			DAVI    
			& {HW} 
			& 0.271  &0.679 &0.302 &0.718 \\
			\hline
		\end{tabular}
	\end{center}
	Note: VI estimates are variational means of calibrated VAs.
\end{table}

 \begin{figure}
	\centering
	\includegraphics[width=1\linewidth]{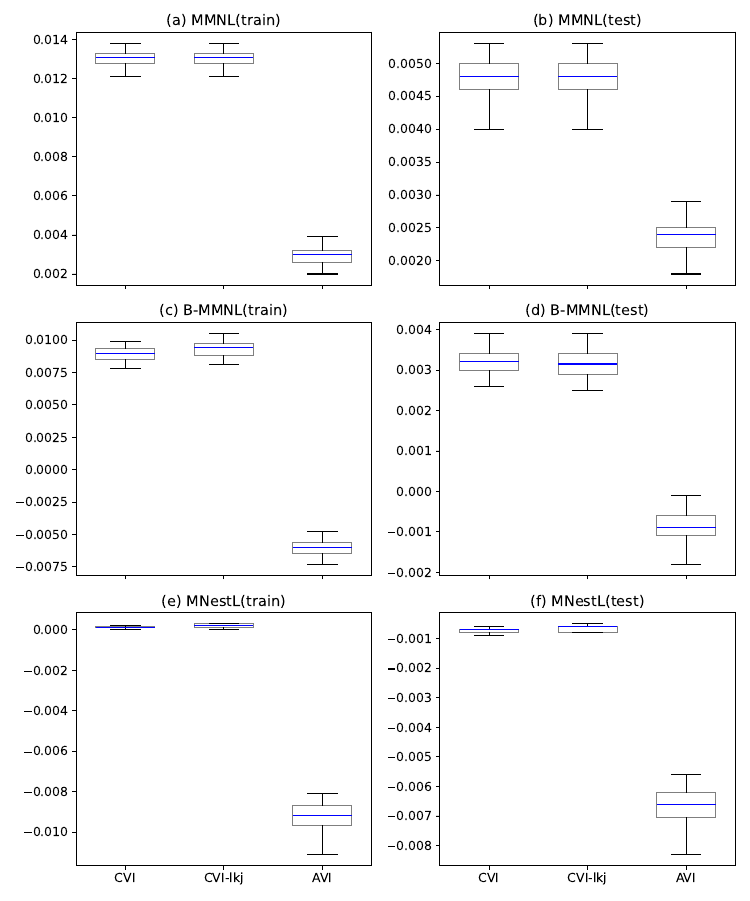}
	\caption{Comparison of log-scores from four VI methods. Values represent differences between each VI method and DAVI. Each row corresponds to one model (MMNL, B-MMNL, MNestL); the first column shows training results and the second column shows testing results. Positive values indicate higher predictive accuracy than DAVI, whereas negative values indicate that DAVI performs better. Boxplots exclude outliers, defined as observations more than 1.5 × IQR from the box.}
	\label{fig:robust_ls}
\end{figure}
\FloatBarrier
\clearpage

% PART D
\setcounter{figure}{0}
\setcounter{table}{0}
\renewcommand{\thetable}{D\arabic{table}}
\renewcommand{\thefigure}{D\arabic{figure}}

\noindent {\bf \Large{Part~D: Additional details on data cleaning of the Carbo-Loading data.}}\\
\ \\
Carbo-Loading dataset\footnote{https://www.dunnhumby.com/source-files/} provides household-level transactions over 2 years for four commodities: pasta, pasta sauce,  syrup and pancake mix. The data consists of three tables. The transactions table contains transactions details, including universal product code (UPC), transaction amount in dollars, household and store identifiers. Each transaction in the table belongs to one ``basket'', which represents one shopping instance, and each basket may have multiple transaction records. The product table contains product details, including UPC, product description, brand and packaging size. The causal table provides promotion activity information for each UPC at each week. For example, ``Feature '' describes the location of the product on the weekly mail, and ``Display'' describes the location of in-store display. The causal table only includes data from week 43. 
\newpage

The detailed steps for data cleaning and imputation for the MMNL and MNestL models are as follows:
\begin{enumerate}
	\item Remove all transaction records before week 43 due to missing promotion data.
	\item  Combine the transactions, product details, and promotion activity tables using UPC, week and store as keys.
	\item Keep only pasta transactions.
	\item Keep transactions for 21 UPCs, representing four popular pasta types (thin spaghetti, regular spaghetti, elbow macaroni, and angel hair) from four brands (Private Label, Barilla, Mueller, and Creamette). Note that different UPCs may belong to the same brand and pasta type due to variations in packaging. This filtering results in 555,911 transactions, accounting for 40.32\% of the total 1,378,686 pasta transactions.
	\item Extract and convert packaging weight measurements into ounces.
	\item For the “Feature” and “Display” variables, set the variables to 1 if there are recorded promotion activities, and 0 otherwise.
	\mycomment{
	\item Retain brands with at least $5\%$ market share within their respective categories. Group smaller brands into “Other Pasta” and “Other Sauce”. The pasta category has 6 major brands with more than $5\%$ market share: private label, Barilla, Creamette, Mueller, private label premium, Ronzoni, and brands with less than $5\%$ but more than $1\%$ market shares are aggregated to ``other pasta''. \footnote{Strictly speaking private label, private label premium and private label value are also collections of different home brands.}. The pasta sauce category has 5 major brands with more than $5\%$ market share: Ragu, Prego, private label, Hunt's, Classico, and brands with less than $5\%$ but more than $1\%$ market shares are aggregated to ``other sauce''. }
	\item Calculate the per-unit price by dividing the total expenditure on the product (UPC) by its packaging weight. 
	\item (a) Use the average price of the same pasta type and brand at the same store on the same day. (b) If the price remains missing, use the average price from the same store during the same week. (c) If no price can be assigned after these steps, label the brand as unavailable for that transaction.
	\item Use steps (a) and (b) above for imputing “Feature” and “Display” variables, but using the maximum value instead of mean. If values remain missing, assign 0 to indicate no promotional activity.
	\item Randomly allocate $80\%$ of transactions from each store to the training dataset and the remaining baskets to the testing data. 
	\item The private label pasta is chosen as the reference alternative because it has the largest market share among all brands and commodities. {For identification purposes, we drop transactions where the reference alternative is not available. The resulting data has 381 groups, 438,774 transaction in the training data, and 109,873 transactions in the testing data.}	
\end{enumerate}
Table \ref{tab:pasta_summary} summarizes the availability and market share of each brand-type combination in the training dataset. Creamette angel hair is not available in all transactions. Availability is calculated as the number of choice occasions where a specific product is available as a percentage of all choice occasions. Market shares are calculated as the number of choice occasions where a specific product is chosen as a percentage of all choice occasions.
\clearpage

\noindent  {\bf {D.2: Carbo-Loading data for the B-MMNL model}}\\
In the MixBC model section, we allow each bundle to contain up to two alternatives. If a bundle includes two alternatives, we impose the restriction that one must be a pasta and the other a pasta sauce. The detailed steps for data cleaning and imputation for the MixBC models are as follows:
\begin{enumerate}
	\item Repeat steps in D.1 to obtain pasta transaction information. Keep transaction IDs of both the training and testing data.
	\item Repeat step 1 and step 2 in D.1, and keep pasta sauce transactions.
	\item Extract and convert packing weight measurements into ounces.
	\item For the ``Feature'' and ``Display'' variables, set the variables to 1 if there are recorded promotion activities, and 0 otherwise.
	\item Keep 4 major pasta sauce brands: ``Ragu'', ``Prego'', ``Private Label'' and ``Hunt's'', aggregate other brands into ``Mixed''. The 4 major brands account for $80\%$ of pasta sauce transactions.
	\item Calculate the per-unit price by dividing the total expenditure on each sauce alternative by total weight in each basket. If there are multiple transaction records of the same sauce alternative in one basket, then keep 1 record for each alternative and drop the duplicates. $958,771$ transactions remain after this step.
	\item There are $74,449$ baskets with more than $1$ sauce alternatives. To ensure at most one sauce alternative in each basket, aggregate multiple brands in one basket as ``Mixed''.
	\item For the missing prices: (a) Use the average price of the same product at the same store on the same day. (b) If the price remains missing, use the average price from the same store during the same week. (c) If no price can be assigned after these steps, label the product as unavailable for that transaction.
	\item Use steps (a) and (b) above for imputing “Feature” and “Display” variables, but using the maximum value instead of mean. If values remain missing, assign 0 to indicate no promotional activity.
	\item Combine the sauce transaction data with the pasta transaction data using ``store'' and ``basket'' as key. One sauce transaction may be mapped to multiple pasta transaction records, as purchases of different pasta alternatives in one basket are separated. 
	\item Fill missing values after combination using means of baskets in the same store and the same day. If still missing, fill missing values using the means of basket in the same store and the same week. Mark as unavailable if still missing.
	\item  Divide the transactions that include pasta into training and testing data set using the transaction IDs from D.1. For the remaining sauce only transactions, randomly allocate $80\%$ of transactions from each store to the training data and the remaining to the testing data.
	\item The private label pasta is chosen as the reference alternative because it has the largest market share among all brands and commodities. {For identification purposes, we also drop transactions where the reference alternative is not available. The resulting data has $381$ groups, $959,050$ transaction in the training data, and $240,192$ transactions in the testing data.}
\end{enumerate}
Table \ref{tab:bundle_summary} summarizes the availability and market share of each bundle in the training dataset. Availability is calculated as the number of choice occasions where a specific bundle is available as a percentage of all choice occasions. When a bundle consists of two products, the bundle is defined as available when both products are available at the choice occasion. Market shares are calculated as the number of choice occasions where a specific bundle is chosen as a percentage of all choice occasions.

\begin{table}[thb!]
	\captionsetup{skip=2pt}
	\centering
	\caption{Availability and Market Share of Pasta and Pasta Sauce }
	\label{tab:bundle_summary}
	\footnotesize
	
	% =======================
	% Panel A: MMNL (4x4 x 2) — Thin×PriLab corrected
	% =======================
	\begin{tabular}{lcccccc}
		\toprule
		& \multicolumn{6}{c}{\textbf{A: Availability (\%)}}                             \\
	&  Ragu   &  Prego  & Private Label &    Hunt's    & Mixed &Singleton Pasta\\
	&         &         &               &      &      & \\
	\textbf{\underline{Thin Spaghetti}} &         &         &               &            & &\\
	Prv-Lab                    & 99.98  &99.06  &97.73  &92.39  &99.33&100\\
	Barilla                        & 54.94  &54.70   &53.78  &51.34  &54.81&54.94\\
	Mueller                     &37.97  &37.81  &37.41  &34.13  &37.82&37.98\\ 
	Creamette 				&41.21  &40.89  &40.31  &40.70   &41.00&41.22\\
	\textbf{\underline{Spaghetti}} &         &         &               &             &&\\
	Prv-Lab                    & 98.13  &97.29  &96.03  &90.85  &97.55&98.14\\
	Barilla 						&57.02  &56.72  &55.89  &53.02  &56.89&57.03\\
	Mueller						&43.65  &43.46  &42.90&   39.40&   43.52&43.66\\
	Creamette				&44.67  &44.32  &43.66  &44.06  &44.41&44.68\\
	\textbf{\underline{Macaroni}} &         &         &               &             &&\\
	Prv-Lab                    &93.86  &93.05  &91.90   &87.22  &93.26&93.87\\
	Barilla 						&36.85  &36.66  &36.19  &33.76  &36.72&36.85\\
	Mueller 					&40.19  &39.93  &39.55  &36.45 & 39.98&40.19\\
	Creamette 				&43.03  &42.62  &42.05  &42.33  &42.73&43.03\\
	\textbf{\underline{Thin Spaghetti}} &         &         &               &     &        &\\
	Prv-Lab              & 87.56  &86.95  &85.71  &81.17  &87.15 &87.57\\
	Barilla					&54.91  &54.65  &53.86  &50.84  &54.84&54.93\\
	Mueller				 &34.75  &34.50   &34.01  &34.29  &34.63&34.76\\
	&         &         &               &     & &\\
	Singleton Sauce&    99.98     &    99.06    &     97.73         &   92.39 & 99.33&--\\
	\bottomrule
	\end{tabular}
	\vspace{2.0ex}
	
	% =======================
	% Panel B: MixMC (category-native labels; first column is PriLab for pasta brand)
	% =======================
	\begin{tabular}{lcccccc}
	& \multicolumn{6}{c}{\textbf{B: Market Shares (\%)}}                             \\
	&  Ragu   &  Prego  & Private Label &    Hunt's    & Mixed &Singleton Pasta\\
&         &         &               &      &      & \\
\textbf{\underline{Thin Spaghetti}} &         &         &               &            & &\\
Prv-Lab                    & 2.456 &1.030 &1.052 &0.748& 0.839&4.225\\
Barilla                        & 0.288  &0.152   &0.040  &0.060  &0.258&0.698\\
Mueller                     &0.352 &0.169 &0.035 &0.044 &0.106&0.421\\ 
Creamette 				&0.404 &0.172  &0.037 &0.087   &0.125&0.613\\
\textbf{\underline{Spaghetti}} &         &         &               &             &&\\
Prv-Lab                    & 2.204  &0.908  &1.086  &0.690  &0.710&3.467\\
Barilla 						&0.321  &0.175  &0.051 &0.066  &0.276&0.695\\
Mueller						&0.505  &0.220 &0.050&   0.069&   0.126&0.503\\
Creamette				&0.535  &0.233  &0.054  &0.130  &0.151&0.711\\
\textbf{\underline{Macaroni}} &         &         &               &             &&\\
Prv-Lab                    &0.543  &0.192 &0.267  &0.176 &0.202&5.495\\
Barilla 						&0.065& 0.030& 0.011& 0.017& 0.052&0.743\\
Mueller 					&0.082& 0.034& 0.010& 0.015& 0.032&1.144\\
Creamette 				&0.139& 0.047& 0.015& 0.036& 0.045&1.387\\
\textbf{\underline{Thin Spaghetti}} &         &         &               &     &        &\\
Prv-Lab              & 0.833& 0.357& 0.349& 0.235& 0.481&1.926\\
Barilla					&0.239& 0.130& 0.037& 0.049& 0.286&0.776\\
Mueller				 &0.217& 0.090& 0.025& 0.049& 0.116&0.435\\
&         &         &               &     & &\\
Singleton Sauce&    20.837     &    8.793   &     6.197         &   4.614 & 13.810&--\\
\bottomrule
\end{tabular}

	\vspace{0.8ex}
	\begin{minipage}{0.85\linewidth}
		\footnotesize
		Panel A reports average availability (in percent) and Panel B reports market shares (in percent). Each cell corresponds to a bundle defined by a pasta (rows) and a sauce (columns). The last column reports singleton pasta bundles (pasta purchased without sauce), while the last row reports singleton sauce bundles (sauce purchased without pasta).
	\end{minipage}
\end{table}

\FloatBarrier
\clearpage

% PART E
\setcounter{figure}{0}
\setcounter{table}{0}
\renewcommand{\thetable}{E\arabic{table}}
\renewcommand{\thefigure}{E\arabic{figure}}

\noindent {\bf \Large{Part~E: Additional details from the consumer choice application.}}\\
\ \\
\noindent  {\bf {E.1: Evaluating price elasticity from calibrated posterior}}\\
Let $Pr(Y_{i}= j \mid \alphavec_i)$ denote the probability that an individual in store $i$ chooses alternative $j$ as a function of random coefficients $\alphavec_i$, and let $\tilde{x}_{2,ij} = \exp({x}_{2,ij})$ represents the price per ounce of alternative $j$ in store $i$\footnote{Here, we omit the subscript $t$ for notational simplicity.}. The own price elasticity of a MNL model \citep[p.~59]{trainDiscreteChoiceMethods2009} is defined as $E_{ij} =\frac{\partial Pr(Y_{i}= j )}{\partial \tilde{x}_{2,ij}} \frac{\tilde{x}_{2,ij}}{Pr(Y_{i}= j )}$, where $Pr(Y_{i}= j )$ denotes the choice probability of alternative $j$ in store $i$, conditional on the model parameters and covariates. The MMNL equivalent can be defined as follows:
\begin{align}
	E_{ij} = \int _{\alpha_i}\frac{\partial Pr(Y_{i}= j \mid \alphavec_i)}{\partial \tilde{x}_{2,ij}} \frac{\tilde{x}_{2,ij}}{Pr(Y_{i}= j \mid \alphavec_i)} d\alphavec_i.\label{equ:elasticity}
\end{align}
See, for example \cite{greeneDoesScaleHeterogeneity2010}, for a similar definition of elasticity. As the integrals can not be computed directly, we use $n_{sim}$ Monte Carlo draws from the fitted VAs of $\alphavec_i$ to obtain an estimate of elasticity:
\begin{align}
	\widehat{E}_{ij} =\frac{ 1}{n_{sim}}\sum_{n_{iter} = 1}^{n_{sim}}\frac{\partial Pr(Y_{i}= j \mid \alphavec^{n_{iter}}_i)}{\partial \tilde{x}_{2,ij}} \frac{\tilde{x}_{2,ij}}{Pr(Y_{i}= j \mid \alphavec^{n_{iter}}_i)}. \label{equ:elasticity_sim}
\end{align}

We follow the “average case” approach \citep{williamsUsingMarginsCommand2012} %, Hensher_2015_othermatter}
by fixing $\tilde{x}_{2,il}$ for all $l\neq j$ at their sample means and varying $\tilde{x}_{2,ij}$ across a grid of values. This allows us to evaluate how the own-price elasticity changes as a function of price. %For further details, see Part B of the Web Appendix.

The following pseudo-code outlines the steps to evaluate predictive elasticity based on calibrated variational posterior. 
\begin{algorithm}[H]
	\caption{Predictive price elasticity}
	\begin{algorithmic}[1]
		\State Generate a grid of $n_{grid}$ evenly spaced values $\tilde{\xvec}^{grid}$ over $10\%$ percentile to $90\%$ percentile of observed price of each alternative.
		\State Compute mean covariate values $X^{mean}$, where $X^{mean}_{j,ind} = \sum_{i = 1}^{S}\sum_{t = 1}^{T_i} x_{i,t,j,ind} / \sum_{i = 1}^{S}T_i$ for $j = 1, \dots,J$ and $ind = 1,\dots,m$. $X^{mean}_{j,2}$ is the mean of prices for alternative $j$. $J$ is the number of alternatives and $m$ is the dimension of $\xvec_{i,t,j}$.
		\State Set discrete covariates in $X^{mean}$ to baseline values ($0$).
		
		\For{$n_{\text{iter}} = 1$ to $n_{\text{sim}}$}
		\State Generate $\psivec^{(n_{iter})} \sim  q_{\lambda^*}(\psivec)$
		\For{$j = 1$ to $J$}
		\For{$n_v = 1$ to $n_{grid}$}
		\State Set $X=X^{mean}$.
		\State Set $X_{j,2}=\xvec^{grid}_{n_v}$.
		\State Evaluate the choice probability $Pr^{n_v}(Y_{i}= j \mid \alphavec^{n_{iter}}_i,X)$.		
		\EndFor
		\State Evaluate $\frac{\partial Pr^{n_v}(Y_{i}= j \mid \alphavec^{n_{iter}}_i)}{\partial \xvec^{grid}_{n_v}}$ for $n_v = 1, \dots, n_{grid}$ and elasticity $E^{n_{iter},n_v}_{i,j}$.
		\EndFor
		\EndFor
		\State Compute mean elasticity over simulations: $\widehat{E}^{n_v}_{i,j} = 1/n_{sim}\sum_{n_{iter} = 1}^{n_{sim}}E^{n_{iter},n_v}_{i,j}$ for $n_v = 1, \dots, n_{grid}$ and $j = 1,\dots,J$.
	\end{algorithmic}
\end{algorithm}

Here we evaluate $\frac{\partial Pr^{n_v}(Y_{i}= j \mid \alphavec^{n_{iter}}_i)}{\partial \xvec^{grid}_{n_v}}$ numerically using the ``gradient'' function in numpy \citep{harrisArrayProgrammingNumPy2020}. \\
\ \\

\noindent  {\bf {E.2: Evaluating probability for the B-MMNL model on the pasta margin}}\\
To make the $\mathbb{F}1$ score from the B-MMNL model comparable to those from the MMNL and MNestL models, we evaluate predictive performance using the {marginal probability of pasta choice}. Specifically, for pasta alternative $j = 1,\dots,15$, we define
\begin{align*}
	Pr(Y_{it,pasta}=j) = \frac{\sum_{b\in \calC(j)}Pr(Y_{it} = b)}{1 - P^0_{it}},
\end{align*}
where $Pr(Y_{it,pasta}=j)$ denotes the marginal probability that pasta $j$ is chosen, $\mathcal C(j) = \{\,  \calB_r : j \in \calB_r \,\}$ denotes the set of bundles that include pasta $j$, and $\Pr(Y_{it} = b)$ is the probability of choosing bundle $b$ predicted by the B-MMNL model. The denominator
\[
P^0_{it} = \sum_{s=16}^{20} \Pr(Y_{it} = s)
\]
represents the predicted probability of choosing a sauce-only singleton bundle, indexed by $s=16,\dots,20$. We evaluate the $\mathbb{F}1$ score using the sub-sample of transactions where a pasta is chosen in the observed data. 

\ \\

\noindent  {\bf {E.3: Additional results from the consumer choice application}}\\
\begin{table}[thb!]
	\captionsetup{skip=2pt}
	\centering
	\caption{Estimated variational means and standard deviations of $\xivec$ corresponding to \textit{lnprice}.}
	\label{tab:coef_lnprice}
	\footnotesize
	
	% =======================
	% Panel A: MMNL (4x4 x 2) — Thin×PriLab corrected
	% =======================
	\begin{tabular}{@{\hspace{5\tabcolsep}}cl@{\hspace{4\tabcolsep}}ccccccc@{\hspace{5\tabcolsep}}}
		\midrule\midrule
		&&      &                        \multicolumn{5}{c}{\textbf{Panel A: MMNL \ \ \ \ \ \ }}         &                     \\
		&&  &     \multicolumn{4}{c}{\textbf{Pasta brand}}       & &   \\ 
		\cline{4-7}
		&&        &        &        &        &          &          &    \\
		&& & PriLab &  Bar   &  Mue   &           Cre            &  &   \\
		&\textbf{Pasta type} & &        &        &        &          &                 &   \\
		&Thin Spaghetti 
		& & --    & -0.510 (0.015)& -0.577(0.018)& -0.626 (0.018)&& \\
		
		&Spaghetti 
		& & -0.542 (0.013)& -0.533 (0.018)& -0.611 (0.022)& -0.683 (0.018)&& \\
		
		&Macaroni 
		& & -0.512 (0.014)&-0.545 (0.018 )& -0.478 (0.018)& -0.588 (0.017)&& \\
		
		&Angel Hair 
		& & -0.489 (0.015)& -0.552 (0.015)& -0.571 (0.020)& --& &\\
		\midrule\midrule
	\end{tabular}
	\vspace{1.0ex}
	
	% =======================
	% Panel C: MixNL (4x4 x 2) — Thin×PriLab corrected
	% =======================
	\begin{tabular}{@{\hspace{5\tabcolsep}}cl@{\hspace{4\tabcolsep}}ccccccc@{\hspace{5\tabcolsep}}}
		&&           &                  \multicolumn{5}{c}{\textbf{Panel B: MNestL \ \ \ \ \ \ }}            &                  \\
		&&   &    \multicolumn{4}{c}{\textbf{Pasta brand}}       & &  \\ \cline{4-7}
		&& &        &        &        &            &               &    \\
		&& & PriLab &  Bar   &  Mue   &           Cre            &   & \\
		&\textbf{Pasta type} & &        &        &        &                          & &   \\
		&Thin Spaghetti
		& & -- & -0.637 (0.018) & -0.719 (0.023) & -0.786 (0.021) & & \\
		
		&Spaghetti
		& & -0.579 (0.016) & -0.593 (0.017) & -0.695 (0.020) & -0.774 (0.018) & & \\
		
		&Macaroni
		& & -0.547 (0.014) & -0.615 (0.026) & -0.550 (0.017) & -0.651 (0.018) & & \\
		
		&Angel Hair
		& & -0.519 (0.016) & -0.612 (0.016) & -0.630 (0.018) & -- & &  \\ \midrule\midrule
	\end{tabular}
	\vspace{1.0ex}
	
	% =======================
	% Panel B: MixMC (category-native labels; first column is PriLab for pasta brand)
	% =======================
	\begin{tabular}{@{\hspace{5\tabcolsep}}cl@{\hspace{4\tabcolsep}}cccccc}
	&&           &                  \multicolumn{5}{c}{\textbf{Panel C: B-MMNL \ \ \ \ \ \ }}                             \\
	&&   &    \multicolumn{4}{c}{\textbf{Pasta brand}}       &  \\ \cline{4-7}
	&& &        &        &        &            &                  \\
	&& & PriLab &  Bar   &  Mue   &           Cre            &   \\
	&\textbf{Pasta type} & &        &        &        &                          &   \\
	&Thin Spaghetti
	& & -- &-1.417(0.013)& -1.468(0.018)& -1.918(0.021)& \\
	
	&Spaghetti
	& & -0.690(0.013)&-1.352(0.014)&-1.427(0.015)& -1.744(0.021)& \\
	
	&Macaroni
	& & -0.900(0.011)&-1.637(0.017)& -1.539(0.016)& -1.822(0.020) &  \\
	
	&Angel Hair
	& & -0.938(0.014)& -1.407(0.014)&-1.869(0.018)& -- &   \\ \midrule
	&&& Ragu & Prego & Prv-Lab & Hunt`s&Mixed\\
	&Sauce Brand &&-0.905(0.012)&-0.943(0.014)&-0.611(0.015)& -1.050(0.016)& -0.659(0.012) \\\bottomrule
	\end{tabular}
	
	\vspace{0.75ex}
	\begin{minipage}{0.98\linewidth}
		\footnotesize
		Panels A and B report the estimates by pasta type (rows) and brand (columns, PriLab = Private Label; Bar = Barilla; Mue = Mueller; Cre = Creamette) for the MMNL and MNestL models, respectively, with the reference alternative and unavailable brand–type combinations (e.g., products not sold for a given brand/type) shown as “–”. Panel C reports the corresponding estimates for the B-MMNL specification: the first four rows report the estimates for different pasta brand-type combinations in the same ways as Panel A, the last row reports the estimates for different pasta sauce brands.  Entries are the fitted Gaussian variational means, with standard deviations given in parentheses.
	\end{minipage}
\end{table}

\begin{table}[htbp]
	\centering
	\caption{Complementary effects between pasta and pasta sauce brands}
	\label{tab:complementary_effects_full}
	\footnotesize
	\begin{tabular}{lccccc}
		\midrule\midrule
		& \multicolumn{5}{c}{\textbf{Pasta sauce brands}} \\ \cline{2-5}
		&         &         &               &             & \\
		&  Ragu   &  Prego  & Private Label &    Hunt`s  & Mixed  \\
		&         &         &               &             \\
		\textbf{\underline{Thin Spaghetti}} &         &         &               &            &  \\
		Prv-Lab                    & -2.1477 &-2.1431 &-1.7780  &-1.8313 &-2.7933\\
		Barilla                        & -2.5435 &-2.3330&  -3.2412& -2.5694 &-2.3077  \\
		Mueller                     &-1.9418& -1.8820  &-3.2062 &-2.2658 &-2.7493   \\ 
		Creamette 				&-2.0614 &-1.9068 &-2.9909 &-2.3104 &-2.6289\\
		\textbf{\underline{Spaghetti}} &         &         &               &       &      \\
		Prv-Lab                    & -2.1123& -2.1313 &-1.6325& -1.7452& -2.7730\\
		Barilla 						&-2.4492& -2.1750&  -3.0794& -2.4949& -2.2326\\
		Mueller						&-1.7350&  -1.7536& -2.9931& -2.0380&  -2.6794\\
		Creamette				&-1.9181 &-1.7422& -2.7698& -2.0298& -2.5619\\
		\textbf{\underline{Macaroni}} &         &         &               &   &          \\
		Prv-Lab                    & -3.9863& -4.1123& -3.4869& -3.6298& -4.4249\\
		Barilla 						&-4.1593& -4.1052& -4.7552& -3.9159& -3.9972\\
		Mueller 					&-4.4428& -4.4159& -5.5271& -4.3820&  -4.8237\\
		Creamette 				&-3.9816& -4.0336& -4.6484& -4.0295& -4.3627\\
		\textbf{\underline{Thin Spaghetti}} &         &         &               &     &        \\
		Prv-Lab              & -2.4869 &-2.4964& -2.1426& -2.2371 &-2.6364\\
		Barilla					&-2.8495 &-2.6242& -3.4670&  -2.8077 &-2.3848\\
		Mueller				 &-2.3448 &-2.2644& -3.0618& -2.4984 &-2.4576\\
		\bottomrule
	\end{tabular}
	\par\vspace{1ex}
	%	\raggedright
	\begin{minipage}{0.8\linewidth}
		\centering
		\footnotesize
		\begin{itemize}[]
			\item The entries represent the estimated $\gamma_r$ for bundle $r$, consisting of the pasta in the given row and the pasta sauce in the given column. 
		\end{itemize}
	\end{minipage}
\end{table}

\clearpage

\end{document}